\documentclass[aps,prc,twocolumn,superscriptaddress,showpacs]{revtex4-2}
\usepackage{lineno,hyperref}
\usepackage{graphicx}% Include figure files
\usepackage{dcolumn}% Align table columns on decimal point
\usepackage{bm}% bold math
\usepackage{amsmath}
\usepackage{amssymb}
\usepackage{amsfonts,times}
\usepackage{latexsym}
\usepackage{siunitx}
\usepackage{soul}

\usepackage[figuresright]{rotating}
\usepackage[version=3]{mhchem}

\def\Panda{\texorpdfstring{$\overline{\mbox{P}}${ANDA}}{Panda}\ }%\xspace}%
\def\panda{\texorpdfstring{$\overline{\mbox{P}}${ANDA}}{Panda}\ }%\xspace}%

\def\kevc1{\ifmmode\mathrm{\ keV/{\mit c}}
          \else$\mathrm{\ keV/{\mit c}}$\fi}

\def\MeVc1{\,MeV/{\mit c}}

\def\mevc1{\ifmmode\mathrm{\ MeV/{\mit c}}
          \else$\mathrm{\ MeV/{\mit c}}$\fi}
\def\gevc1{\ifmmode\mathrm{\ GeV/{\mit c}}
          \else$\mathrm{\ GeV/{\mit c}}$\fi}
\def\GeVc1{\ifmmode\mathrm{\ GeV/{\mit c}}
          \else$\mathrm{\ GeV/{\mit c}}$\fi}
\def\kevc2{\ifmmode\mathrm{\ keV/{\mit c}^2}
          \else$\mathrm{\ keV/{\mit c}^2}$\fi}
\def\Mevc2{\ifmmode\mathrm{\ MeV/{\mit c}^2}
          \else$\mathrm{\ MeV/{\mit c}^2}$\fi}
\def\Gevc2{\ifmmode\mathrm{\ GeV/{\mit c}^2}
          \else$\mathrm{\ GeV/{\mit c}^2}$\fi}
\def\Gev2c2{\ifmmode\mathrm{\ GeV^2/{\mit c}^2}
          \else$\mathrm{\ GeV^2/{\mit c}^2}$\fi}
\def\Pgp{\ifmmode\math{p}
         \else$\mathrm{p}$\fi}
\def\Pagp{\ifmmode\mathrm{\overline{p}}
         \else$\mathrm{\overline{p}}$\fi}
\def\Pgn{\ifmmode\mathrm{n}
         \else$\mathrm{n}$\fi}
\def\Pagpn{\ifmmode\mathrm{\overline{n}}
         \else$\mathrm{\overline{n}}$\fi}
\def\Pp{\ifmmode\mathrm{p}
         \else$\mathrm{p}$\fi}
\def\Pap{\ifmmode\mathrm{\overline{p}}
         \else$\mathrm{\overline{p}}$\fi}

\def\Pn{\ifmmode\mathrm{n}
         \else$\mathrm{n}$\fi}
\def\Pan{\ifmmode\mathrm{\overline{n}}
         \else$\mathrm{\overline{p}}$\fi}
\def\Py{\ifmmode\mathrm{Y}
         \else$\mathrm{Y}$\fi}
\def\Pay{\ifmmode\mathrm{\overline{Y}}
         \else$\mathrm{\overline{Y}}$\fi}

\def\PgL{\ifmmode\mathrm{\Lambda }
          \else$\mathrm{\Lambda }$\fi}
\def\PagL{\ifmmode\mathrm{\overline{\Lambda }}
            \else$\mathrm{\overline{\Lambda }}$\fi}
\def\PgS{\ifmmode\mathrm{\Sigma }
          \else$\mathrm{\Sigma }$\fi}
\def\PagS{\ifmmode\mathrm{\overline{\Sigma }}
            \else$\mathrm{\overline{\Sigma }}$\fi}
\def\PgSp{\ifmmode\mathrm{\Sigma^+}
          \else$\mathrm{\Sigma^+}$\fi}
\def\PagSp{\ifmmode\mathrm{\overline{\Sigma^+}}
            \else$\mathrm{\overline{\Sigma^+}}$\fi}
\def\PgSm{\ifmmode\mathrm{\Sigma^-}
          \else$\mathrm{\Sigma^-}$\fi}
\def\PagSm{\ifmmode\mathrm{\overline{\Sigma^-}}
            \else$\mathrm{\overline{\Sigma^-}}$\fi}
\def\PgSn{\ifmmode\mathrm{{\Sigma }^0}
            \else$\mathrm{{\Sigma }^0}$\fi}
\def\PagSn{\ifmmode\mathrm{\overline{\Sigma }^0}
            \else$\mathrm{\overline{\Sigma }^0}$\fi}
\def\PgX{\ifmmode\mathrm{\Xi }
          \else$\mathrm{\Xi }$\fi}
\def\PagX{\ifmmode\mathrm{\overline{\Xi }}
            \else$\mathrm{\overline{\Xi }}$\fi}
\def\PgXm{\ifmmode\mathrm{\Xi^-}
          \else$\mathrm{\Xi^-}$\fi}
\def\PagXm{\ifmmode\mathrm{\overline{\Xi^-}}
            \else$\mathrm{\overline{\Xi^-}}$\fi}
\def\PagXp{\ifmmode\mathrm{\overline{\Xi }^+}
            \else$\mathrm{\overline{\Xi }^+}$\fi}

\def\PgOm{\ifmmode\mathrm{\Omega^-}
          \else$\mathrm{\Omega^-}$\fi}
\def\PagOm{\ifmmode\mathrm{\overline{\Omega^-}}
            \else$\mathrm{\overline{\Omega^-}}$\fi}
\def\PgOp{\ifmmode\mathrm{\Omega^+}
          \else$\mathrm{\Omega^+}$\fi}
\def\PagOp{\ifmmode\mathrm{\overline{\Omega }^+}
            \else$\mathrm{\overline{\Omega }^+}$\fi}

\def\PgLc{\ifmmode\mathrm{\Lambda_c}
          \else$\mathrm{\Lambda_c}$\fi}
\def\PagLc{\ifmmode\mathrm{\overline{\Lambda }_c}
            \else$\mathrm{\overline{\Lambda }_c}$\fi}

\def\PgD{\ifmmode\mathrm{D}
          \else$\mathrm{D}$\fi}
\def\PagD{\ifmmode\mathrm{\overline{D}}
            \else$\mathrm{\overline{D}}$\fi}

\def\PgPi{\ifmmode\mathrm{\pi }
          \else$\mathrm{\pi }$\fi}
\def\PagPi{\ifmmode\mathrm{\overline{\pi }}
            \else$\mathrm{\overline{\pi }}$\fi}
\begin{document}

%\preprint{}

\title{Probing small neutron skin variations in isotope pairs by hyperon-antihyperon production in
antiproton--nucleus interactions}


\author{Falk Schupp}
%\thanks{}
\affiliation{Helmholtz-Institut Mainz, Johannes Gutenberg-Universit\"at Mainz, 55099 Mainz, Germany}
\author{Josef Pochodzalla}
\email[Contact author:]{pochodza@uni-mainz.de}
\affiliation{Helmholtz-Institut Mainz, Johannes Gutenberg-Universit\"at Mainz, 55099 Mainz, Germany}
\affiliation{Institute f\"ur Kernphysik, Johannes Gutenberg-Universit\"at Mainz, 55099 Mainz, Germany}
\affiliation{PRISMA$^+$ Cluster of Excellence, Johannes Gutenberg-Universit\"at Mainz, 55099 Mainz, Germany}
\author{Michael~B\"olting}
\affiliation{Helmholtz-Institut Mainz, Johannes Gutenberg-Universit\"at Mainz, 55099 Mainz, Germany}
\author{Martin Christiansen}
\affiliation{Helmholtz-Institut Mainz, Johannes Gutenberg-Universit\"at Mainz, 55099 Mainz, Germany}
\author{Theodoros Gaitanos}
\affiliation{Department of Theoretical Physics, Aristotle University of Thessaloniki, GR-54124 Thessaloniki, Greece}
\author{Horst Lenske}
\affiliation{Institut für Theoretische Physik, Justus-Liebig-Universität Gie\ss en, D-35392 Gie\ss en, Germany}
\author{Marcell~Steinen}
\affiliation{Helmholtz-Institut Mainz, Johannes Gutenberg-Universit\"at Mainz, 55099 Mainz, Germany}


\date{\today}

\begin{abstract}
We propose a new method to study the evolution of the neutron periphery
between different isotopes by considering antiproton--nucleus interactions close to the production threshold of $\Lambda \overline{\Lambda }$ and $\Sigma^-\overline{\Lambda }$ pairs. At low energies, $\Lambda \overline{\Lambda }$ pairs are produced in $\overline{\text{p}} +\text{p}$ collisions, while $\Sigma^-\overline{\Lambda }$ pairs can only be produced in $\overline{\text{p}} +\text{n}$ interactions.
Within a simple geometrical picture we show that the double ratio for the production of $\Sigma^-\overline{\Lambda }$ and $\Lambda \overline{\Lambda }$ pairs for two different isotopes are related to the variation of the neutron skin thickness between the two nuclei.
Performing high statistics calculations with the Gie\ss en Boltzmann--Uehling--Uhlenbeck (GiBUU) transport model for several isotope pairs covering a wide range of elements we verify a strong correlation between the double ratio from the full transport simulations and the schematic model.
This correlation enables us to quantify the potential of the proposed method for precise studies of neutron skin variations in isotope chains.
\end{abstract}
\maketitle
\section{Introduction\label{sec:intro}}
To describe neutron rich matter which eventually appears in dense stellar objects, knowledge on the isospin part of the EoS is indispensable. Luckily, the isospin dependence of the EoS correlates strongly with the distribution of neutrons in nuclei, see \cite{PhysRevC.86.015803,BALDO2016203,doi:10.1063/PT.3.4247} and references therein.
Thus the thickness of neutron skins in nuclei impacts our knowledge on the structure of neutron stars %\cite{PhysRevLett.85.5296,
\cite{PhysRevLett.85.5296,PhysRevLett.86.5647,PhysRevC.64.027302,PhysRevC.64.062802,PhysRevC.79.057301,PhysRevC.87.034301,PhysRevC.93.051303,PhysRevLett.102.122502}.
Furthermore, the evolution of the neutron skin and proton distributions along isotope chains provide important information for our understanding of the nuclear structure over the whole nuclear chart
\cite{PhysRevLett.75.3241,ALKHAZOV1976443,PhysRevC.29.182,PhysRevC.46.1825,PhysRevC.65.014306,PhysRevC.77.034315,PhysRevC.82.044611}.

The neutron skin thickness, ${\Delta }$R$_\text{np}$, is usually defined as the difference
between the root-mean-squared (rms) point radii of the neutron and proton density distributions.
However, this definition is not unique because of possibly significant density variations both, in the nuclear periphery and in the center \cite{PhysRevC.61.044326} (see the discussion in \autoref{sec:sensitivity}).
Since quantal shell effects also influence the nucleon distributions, the focus
of their studies often lies on neutron rich doubly magic nuclei, like $^{40}$Ca, $^{48}$Ca and $^{208}$Pb.

% ----------------------------------------------
\begin{figure*}[tb]
  \centering
  \includegraphics[width=1.0\textwidth]{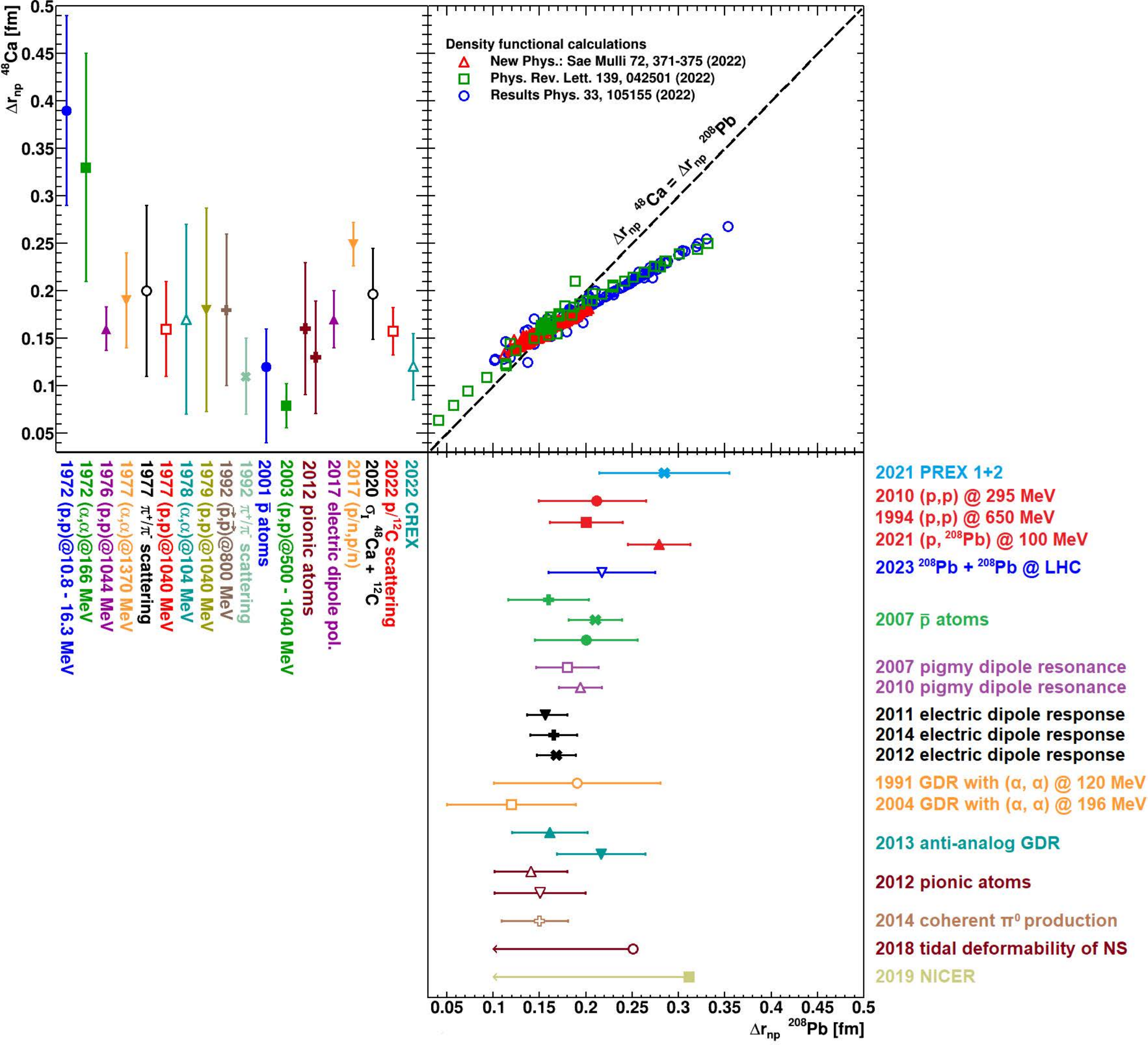}
  \caption{Experimental values for the neutron skin thickness of the doubly magic nuclei $^{48}$Ca (left panel) and $^{208}$Pb (lower right panel). Detailed information on the displayed data are listed in \autoref{tab:nskinCa} and
  \autoref{tab:nskinPb}. The upper right panel shows the relation between the neutron skin thickness of $^{48}$Ca or $^{208}$Pb predicted by various models with different EoS \cite{Horowitz2014,TAGAMI2022105155,Hyun2022}.
  Note, that the combined PREX measurement of $^{208}$Pb \cite{PhysRevLett.108.112502,PhysRevLett.126.172502} and the CREX result for $^{48}$Ca \cite{PhysRevLett.129.042501} are in tension with these model predictions. References are listed in the Appendix.
  }
  \label{fig:CaPb}
\end{figure*}
% ----------------------------------------------

Experimentally charge distributions, which mainly reflects the proton distribution, can be explored by electromagnetic probes like electrons or muons rather precisely. On the other hand, accurate information on neutron radii and neutron skins is scarce.
Experimental techniques to determine the neutron distributions of nuclei using strongly interacting probes range from proton elastic
\cite{BRISSAUD1979141,PhysRevC.65.044306,PhysRevC.67.054605,TAGAMI2022105155,SAKAGUCHI20171}
and inelastic
\cite{Abdul_Magead_2020,PhysRevC.104.024606}
reactions, measurement of interaction cross section of heavy nuclei \cite{PhysRevLett.75.3241,PhysRevC.77.034315,PhysRevLett.124.102501,TAGAMI2022105155,PhysRevC.105.014626,PhysRevLett.131.202302} and antiproton interactions \cite{LENSKE200782}.
The electric dipole response can be studied by
$\alpha$ particle scattering exciting the giant dipole resonance \cite{PhysRevLett.66.1287,KRASZNAHORKAY2004224},
the excitation of the pigmy dipole resonance \cite{PhysRevC.81.041301} and
spin-dipole resonance  \cite{PhysRevLett.82.3216,KRASZNAHORKAY2004224}, the
electric dipole polarizability \cite{PhysRevLett.107.062502,PhysRevLett.118.252501}, pion scattering \cite{PhysRevC.46.1825}, coherent pion photoproduction \cite{PhysRevLett.112.242502},
antiprotonic x-rays \cite{PhysRevLett.87.082501,PhysRevC.65.014306}.
The parity-violating electron scattering asymmetry provides a measure of the
weak charge distribution and hence of neutron distribution in appropriate nuclei \cite{Horowitz2014}.
Finally, since neutron star properties and the neutron skins of nuclei are highly related \cite{doi:10.1063/PT.3.4247}, also astrophysical observations of neutron stars provide in an indirect way constraints on the neutron skin of e.g. $^{208}$Pb \cite{PhysRevLett.120.172702,PhysRevLett.119.161101short,PhysRevLett.126.172503}.
Despite these many experimental opportunities, the neutron distributions still carry considerable systematic uncertainties, see, for instance the discussions in Refs.~\cite{PhysRevC.79.057301,KRASZNAHORKAY2004224,PhysRevC.100.044608}.

Calcium isotopes are of particular value, since for such medium heavy nuclei, already accurate {\em ab initio} calculations are possible \cite{PhysRevC.91.051301,PhysRevLett.127.072501,HEBELER20211}, although in many cases
a quantitative reproduction of nuclear radii is still an open issue \cite{PhysRevC.97.021303,PhysRevC.100.034310}.
Calculations using a wide variety of EoS's show a remarkable correlation between the slope parameter of the symmetry energy
and the neutron skin thickness of $^{48}$Ca or $^{208}$Pb. As consequence there exists a rather model independent correlation between the neutron skin thickness of these two nuclei \cite{Horowitz2014,TAGAMI2022105155,Hyun2022} (see also in the upper right panel of \autoref{fig:CaPb}).
Data available for the two doubly magic nuclei $^{48}$Ca and $^{208}$Pb are exemplified in the left and lower panel in \autoref{fig:CaPb}, respectively. Many of these analyses have sizable systematic uncertainties since the analyses are model dependent
(for a recent analysis see e.g. \cite{PhysRevC.101.044303,PhysRevLett.125.102501}).
In this figure, statistical and systematic uncertainties (if available) were added in quadrature. Detailed information on the displayed data are listed in \autoref{tab:nskinCa} and \autoref{tab:nskinPb} in the Appendix.

A rather clean measure of the neutron distribution with small systematic uncertainties is expected from the electroweak asymmetry in elastic electron-nucleus scattering \cite{Horowitz2014}.
However, the PREX measurement of $^{208}$Pb \cite{PhysRevLett.108.112502,PhysRevLett.126.172502} and the CREX result for $^{48}$Ca \cite{PhysRevLett.129.042501} are in tension with the expected correlation between the neutron skin of $^{48}$Ca and $^{208}$Pb shown in the upper right panel of \autoref{fig:CaPb}. Indeed, for example a recent nonlocal dispersive optical model analysis suggests a neutron skin for $^{208}$Pb %of 0.25$\pm$ 0.05\,fm \cite{PhysRevC.101.044303}. The latest analysis by the same group suggest a neutron skin
of 0.18$^{+0.07}_{-0.06}$\,fm \cite{PhysRevLett.125.102501}, which is significantly lower than the PREX value, and for $^{48}$Ca a skin of 0.22$^{+0.02}_{-0.03}$\,fm, which in turn is significantly larger than the CREX result.
This situation clearly calls for further detailed theoretical \cite{PhysRevLett.129.232501} as well as improved experimental studies.

%------------------------------------------------------

Throughout this work we  present a novel method to explore the variation of the neutron skin between two different isotopes of a given element with high precision and accuracy. For that purpose we consider antiproton--nucleus interactions close to the thresholds of $\Lambda \overline{\Lambda }$ and $\Sigma^-\overline{\Lambda }$ pair production. At low energies, $\Lambda \overline{\Lambda }$ pairs are produced in $\overline{\text{p}}$+p collisions, while $\Sigma^-\overline{\Lambda }$ pairs can only be produced in $\overline{\text{p}}$+n interactions. Unlike other probes, antiprotons are strongly absorbed in the nuclear periphery and, therefore, are particularly sensitive to small variations of the nuclear skin.

Measuring the probabilities p$_{\Lambda\overline{\Lambda }}$ and p$_{\Sigma^-\overline{\Lambda }}$ for the two processes for a reference isotope (I) and a second isotope (II), allows to determine the double ratio
\begin{equation}
  DR = \frac{p^{II}_{\Sigma^-\overline{\Lambda }}\Big/p^{II}_{\Lambda \overline{\Lambda }} }{p^{I}_{\Sigma^-\overline{\Lambda }}\Big/p^{I}_{\Lambda \overline{\Lambda }}}.
\label{eq:dr}
\end{equation}
Such a double ratio does not require the absolute determination of cross sections
and also the different energy dependence for the $\Lambda \overline{\Lambda }$ and $\Sigma^-\overline{\Lambda }$ channels is eliminated. Furthermore, measuring both production probabilities for a given isotope simultaneously, many experimental uncertainties cancel or can be significantly reduced.

\begin{figure}
  \centering
  \includegraphics[width=1.0\columnwidth]{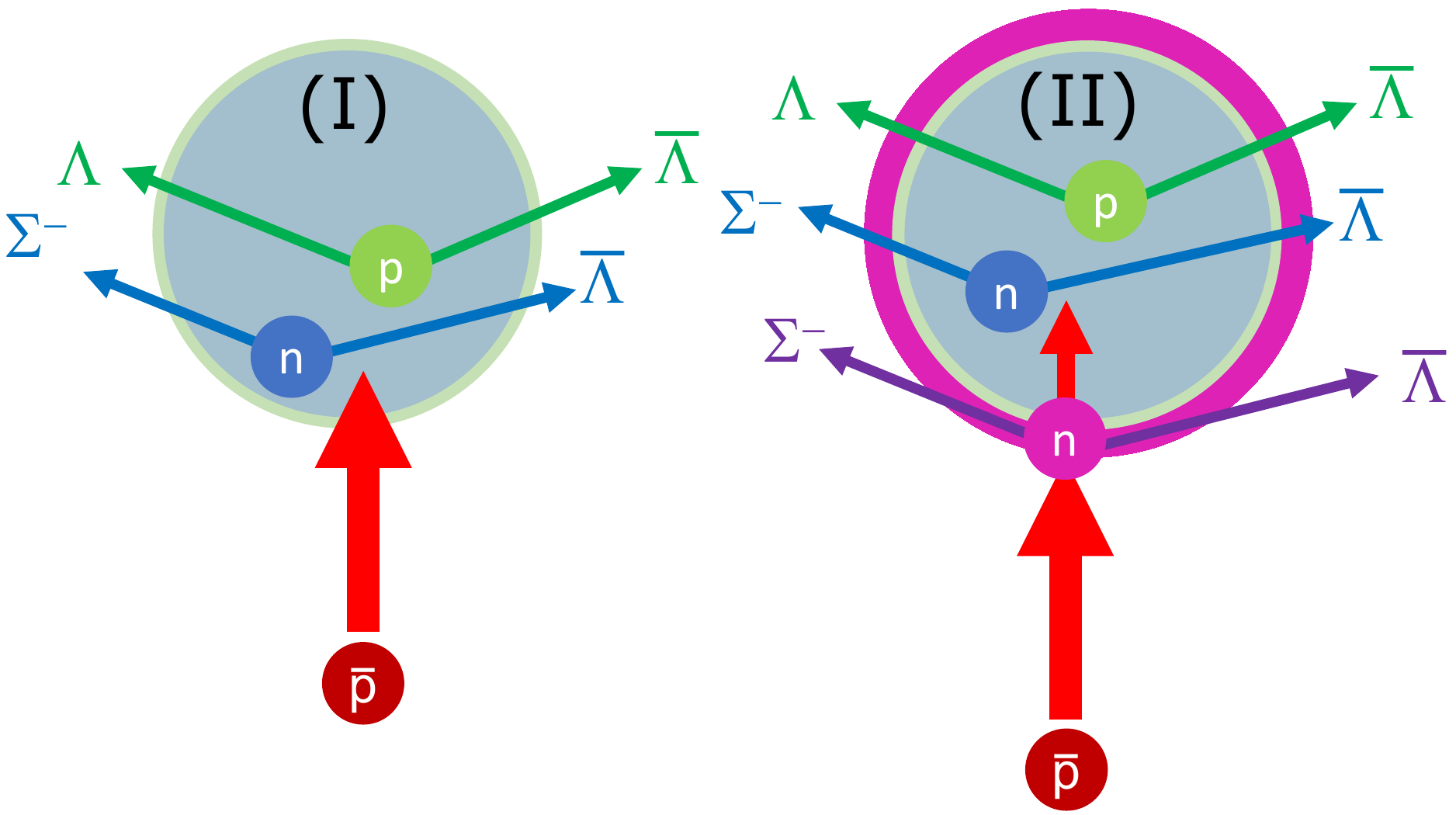}
  \caption{Illustration of the production of $\Lambda \overline{\Lambda }$ and $\Sigma^-\overline{\Lambda }$ pairs in antiproton-nucleus interactions. %While $\Sigma^-\overline{\Lambda }$ pairs can only be produced in $\overline{p}$+n interactions.
  We considering two isotopes (I) and (II) which differ by an additional outer neutron layer with thickness $\Delta_{n}$, shown in pink. The production yields for the two different hyperon-antihyperon pairs provide a measurement of the effective thickness of this additional neutron layer.
  }
  \label{fig:NeuS_scheme}
\end{figure}

Within a simple geometrical picture for central collisions we first show in \autoref{sec_01}
that this ratio is strongly related to the difference of the neutron skin thicknesses of the two considered isotopes (I) and (II) depicted in \autoref{fig:NeuS_scheme}.

In \autoref{sec_02} we present a schematic scenario to describe the full impact parameter range. In \autoref{sec_03}, we compare these calculations with predictions of the Gie\ss en Boltzmann--Uehling--Uhlenbeck (GiBUU) transport model \cite{Buss20121} for various isotope pairs. Using the neutron distributions which describe the initial state in these GiBUU simulations, we find a remarkable correlation between double ratio predicted by the simple picture of \autoref{sec_02} and the double ratio predicted by the complete GiBUU simulations. As a consequence, the simple scenario presented in \autoref{sec_02} enables us to explore in \autoref{sec:sensitivity} the sensitivity of the double ratio to variations of the neutron distributions. As a possible application at the \Panda experiment, we perform a systematic study for the case of $^{40}$Ca and $^{48}$Ca.

% ----------------------------------------------

\begin{table*}[tb]
\begin{ruledtabular}
\centering
\begin{tabular}{llllllll}
%\hline
{\bf isotope}&{\bf abundance}& {\bf exp. charge }&{\bf proton}&{\bf neutron}&  {\bf neutron} & {\bf skin} & {\bf RMF}\\
 & & {\bf radius}&{\bf radius}&{\bf radius}&  {\bf skin} & {\bf difference} & {\bf model} \\
 &{\bf  [\%]}  &{\bf R$_p^{exp}$ [fm]}     &{\bf R$_p$ [fm]}   &{\bf R$_n$ [fm]}    &  {\bf ${\Delta }$R$_{pn}$ [fm]} & {\bf $\Delta_n$ [fm]}& \\

\hline
$^{20}$Ne  & 90.5 & 2.992$\pm$0.008 \cite{DEVRIES1987495}& 2.782 & 2.758 & -0.024 & --    & NL3 \cite{PhysRevC.55.540}  \\
$^{22}$Ne  & 9.3  & 2.986$\pm$0.021 \cite{DEVRIES1987495}& 2.800 & 2.887 &  0.087 & 0.111 & NL3 \cite{PhysRevC.55.540}\\
\hline
$^{40}$Ca  & 96.9  & 3.4776$\pm$0.0019 \cite{ANGELI201369}                & 3.452 & 3.416 & -0.036 &       & NL1 \cite{PhysRevC.55.540}\\
$^{48}$Ca  & 0.187 & 3.4786$\pm$0.0106 \cite{ANGELI201369,GarciaRuiz2016} & 3.525 & 3.731 & 0.206  &0.242  & NL1 \cite{PhysRevC.55.540}\\
$^{40}$Ca  & 96.9  & 3.4776$\pm$0.0019 \cite{ANGELI201369}&                3.391 &3.354 & -0.037 & --     & NL3 \cite{PhysRevC.55.540} \\
$^{48}$Ca  & 0.187 & 3.4786$\pm$0.0106 \cite{ANGELI201369,GarciaRuiz2016}& 3.472 & 3.659 &  0.187 & 0.224 & NL3 \cite{PhysRevC.55.540}\\
$^{40}$Ca  & 96.9  & 3.4776$\pm$0.0019 \cite{ANGELI201369}&                3.396 &3.360 & -0.036 & --     & NL3* \cite{LALAZISSIS200936}\\
$^{48}$Ca  & 0.187 & 3.4786$\pm$0.0106 \cite{ANGELI201369,GarciaRuiz2016}& 3.475 & 3.666 &  0.191 & 0.227 & NL3* \cite{LALAZISSIS200936}\\
%$^{40}$Ca  & 96.9  & 3.4776$\pm$0.0019 \cite{ANGELI201369}&                3.117 &3.090 & -0.028 & --     & Set I \cite{PhysRevC.65.045201}\\
%$^{48}$Ca  & 0.187 & 3.4786$\pm$0.0106 \cite{ANGELI201369,GarciaRuiz2016}& 3.243 & 3.357 &  0.1114 & 0.142 &Set I \cite{PhysRevC.65.045201}\\
\hline
$^{58}$Ni  & 68.1  & 3.770$\pm$0.002 \cite{PhysRevLett.128.022502}& 3.769 &3.768 &  0.000 & --    & NL3 \cite{PhysRevC.55.540}  \\
$^{64}$Ni  & 0.926 & 3.854$\pm$0.002 \cite{PhysRevLett.128.022502}& 3.822 &3.947 &  0.125 & 0.125 & NL3 \cite{PhysRevC.55.540}\\
\hline
$^{129}$Xe  & 26.4 & 4.7775$\pm$0.0050 \cite{ANGELI201369}& 4.768 & 4.932 & 0.164 & --    & NL3 \cite{PhysRevC.55.540}\\
$^{130}$Xe  &4.1   & 4.7818$\pm$0.0049 \cite{ANGELI201369}& 4.776 & 4.950 & 0.174 & 0.010 & NL3 \cite{PhysRevC.55.540}\\
$^{131}$Xe  &21.2  & 4.7808$\pm$0.0049 \cite{ANGELI201369}& 4.784 & 4.968 & 0.184 & 0.020 & NL3 \cite{PhysRevC.55.540}\\
$^{132}$Xe  &26.9  & 4.7859$\pm$0.0048 \cite{ANGELI201369}& 4.792 & 4.986 & 0.194 & 0.030 & NL3 \cite{PhysRevC.55.540}\\
$^{134}$Xe  &10.4  & 4.7899$\pm$0.0047 \cite{ANGELI201369}& 4.809 & 5.023 & 0.213 & 0.049 & NL3 \cite{PhysRevC.55.540}\\
$^{136}$Xe  &8.9   & 4.7964$\pm$0.0047 \cite{ANGELI201369}& 4.826 & 5.059 & 0.233 & 0.069 & NL3 \cite{PhysRevC.55.540}\\
%\hline
\end{tabular}
\caption{Isotopes studied in this work. The second and third column give the natural abundances and the experimental charge radii.
Proton and neutron rms radii and their differences predicted by the relativistic mean field model of Ref.~\cite{PhysRevC.55.540} for stable or long-lived isotopes are listed in the column 4 to 7. The radial distributions are generated during the initialization of the Gie\ss en Boltzmann--Uehling--Uhlenbeck (GiBUU) transport model simulations \cite{Buss20121}.
}
\end{ruledtabular}
\label{tab:radii}
\end{table*}
%------------------------------------------------------------------------------

\section{Motivation of the method: a simple scenario}
\label{sec_01}

Because of the strong absorption of antiprotons in nuclei the production of hyperon--antihyperon pairs happens in the nuclear periphery. For simplicity we consider a nearly central antiproton--nucleus collision (see left part of \autoref{fig:NeuS_scheme}). Close to threshold, the probability $p^{I}_{\Lambda \overline{\Lambda }}$ to produce a $\Lambda \overline{\Lambda }$ pair within the reference nucleus (I) can be written as
\begin{equation}
  p^{I}_{\Lambda \overline{\Lambda }}=\kappa_{\Lambda \overline{\Lambda }}        \cdot \frac{\rho (p)}{\rho (p)+\rho (n)}        \cdot \frac{\sigma_{\Lambda \overline{\Lambda }}}{\sigma_\text{tot}}.
\end{equation}
Here $\sigma_\text{tot}$  denotes the total ${\overline{\text{p}}+A}$ cross section
and $\sigma_{\Lambda \overline{\Lambda }}$ is the elementary $\overline{\text{p}}+{\text{p}} \rightarrow \Lambda \overline{\Lambda }$ cross section.
$\rho (p)$ and $\rho (n)$ denote the %effective
densities in the periphery of the target nucleus of protons and neutrons, respectively.

The factor $\kappa_{\Lambda \overline{\Lambda }}$ describes the loss of $\Lambda \overline{\Lambda }$ pairs due to absorptive re-scattering.
%by the absorption of the produced $\Lambda$ or $\overline{\Lambda }$ after their production.
Similarly, the production probability of $\Sigma^-\overline{\Lambda }$ pairs can be approximated by
\begin{equation}
  p^{I}_{\Sigma^-\overline{\Lambda }}=\kappa_{\Sigma^-\overline{\Lambda }}       \cdot \frac{\rho (n)}{\rho (p)+\rho (n)}                \cdot \frac{\sigma_{\Sigma^-\overline{\Lambda }}}{\sigma_\text{tot}}.
\label{eq:siglamI}
\end{equation}
Because of the large annihilation cross section of antibaryons in nuclei, both, $\kappa_{\Lambda \overline{\Lambda }}$ and $\kappa_{\Sigma^-\overline{\Lambda }}$ are dominated by the $\overline{\Lambda }$ absorption. Therefore, we assume
$\kappa_{\Lambda \overline{\Lambda }} \approx \kappa_{\Sigma^- \overline{\Lambda }} \equiv \kappa_I$.

We now turn to a isotope (II) with a larger neutron number and hence a more extended neutron distribution. For simplicity, we assume that the proton distribution remains identical to the one of nucleus (I) and that the neutron distribution is only extended by an additional skin $\Delta_\text{n}$ at the surface, see pink area in the right part of \autoref{fig:NeuS_scheme}. Such a situation is approximately found in isotope chains of heavy nuclei.
In such a scenario, the production of $\Lambda \overline{\Lambda }$ pairs is reduced by the absorption probability $p_\text{abs}$ of the incident antiprotons within this additional neutron skin $\Delta_\text{n}$:
\begin{equation}
  p^{II}_{\Lambda \overline{\Lambda }}=(1-p_\text{abs})       \cdot \kappa_{II}       \cdot \frac{\rho (p)}{\rho (p)+\rho (n)} \cdot \frac{\sigma_{\Lambda \overline{\Lambda }}}{\sigma_\text{tot}}.
\label{eq:pll2}
\end{equation}
The absorption probability $p_\text{abs}$ can be expressed in terms of the total $\overline{\text{p}}+\text{n}$ reaction cross section
$\sigma_{\overline{\text{p}} \text{n}}$~\cite{Workman:2022ynf} and the integrated skin density $\int_{\Delta_\text{n}} \rho_\text{n}\, \text{d}r_\text{n}$ of the additional neutron skin of nucleus (II) with respect to the reference nucleus (I):
\begin{equation}
  1-p_\text{abs}\approx \exp \Big\{ -\sigma_{\overline{\text{p}}\text{n}}        \cdot \int_{\Delta_\text{n}} \rho_\text{n} \text{d}r_\text{n} \Big\}
\label{eq:pabs}
\end{equation}
Like the incoming antiprotons, the produced $\overline{\Lambda }$ are also absorbed in the additional neutron layer. We use the factorisation ansatz
$\kappa_{II}=\kappa_{I}\cdot \kappa_{n}$.
With this simplification we can express \autoref{eq:pll2} as
\begin{equation}
  p^{II}_{\Lambda \overline{\Lambda }} = \kappa_{n} \cdot (1-p_{abs}) \cdot p^{I}_{\Lambda \overline{\Lambda }}.
\label{eqLLII}
\end{equation}
The production of $\Sigma^-\overline{\Lambda }$ pairs gains an additional component from the
additional neutron skin. On the other hand, the contribution from the inner part of the nucleus (II)
is reduced by the loss of antiprotons in the additional neutron layer:
\begin{multline}
  p^{II}_{\Sigma^-\overline{\Lambda }} = \kappa_{II}                \cdot {p_\text{abs}} \cdot {\frac{\sigma_{\Sigma^-\overline{\Lambda }}}{\sigma_\text{tot}}} \\
\\
+ (1-p_\text{abs}) \cdot \kappa_{II}\frac{\rho (n)}{\rho (p)+\rho (n)} \cdot \frac{\sigma_{\Sigma^-\overline{\Lambda }}}{\sigma_\text{tot}}.
\end{multline}
Here, we assume that the loss of outgoing $\Sigma^-$ and$/$or $\overline{\Lambda }$
for pairs produced in the additional neutron skin is the same as for pairs produced within the core nucleus. We thus obtain for the double ratio
\begin{equation}
  DR = \frac{p^{II}_{\Sigma^-\overline{\Lambda }}\Big/p^{II}_{\Lambda \overline{\Lambda }}}{p^{I}_{\Sigma^-\overline{\Lambda }}\Big/p^{I}_{\Lambda \overline{\Lambda }}}
  = \frac{p_\text{abs} \cdot \frac{\rho (p)+\rho (n)}{\rho (n)} + \Big(1-p_\text{abs}\Big) }{1-p_\text{abs}}
\end{equation}
With the simplifying assumption $\rho (p)$ = $\frac{Z}{N}\cdot\rho (n)$, where Z denotes the element number and N the neutron number of the reference isotope I, we finally find for the double ratio the expression
\begin{equation}
  DR = \frac{1+p_\text{abs}{\cdot}Z/N}{1-p_\text{abs}}
\label{eq:dr01}
\end{equation}
%Note, that this expression
Since the variation of the neutron skin thickness is rather small, the additional absorption probability $p_\text{abs}$ is also small. In this case, we can expand \autoref{eq:dr01} and obtain
a linear relation between the double ratio DR and the absorption probability of the incident antiprotons within the increased neutron skin $\Delta_\text{n}$.
\begin{equation}
DR \approx 1+(1+\frac{Z}{N}) \cdot p_\text{abs}.
\label{eq:dr01b}
\end{equation}
This expression is nearly independent of the considered nuclei and it signals that DR is a direct measure of the increment of (integrated) neutron skin thickness. The only quantity to be known is the total $\overline{\text{p}}+\text{n}$ reaction cross section $\sigma_{\overline{\text{p}} \text{n}}$ which can be determined from experiment by comparing e.g.
${\overline{\text{p}} \text{p}}$ and ${\overline{\text{p}} \text{d}}$ interactions.

\section{Finite impact parameter range \label{sec_02}}

Of course, this simplified geometrical picture has several deficiencies:
\begin{itemize}
\item
In $\Pap$-A collisions one can not constrain the impact parameter and one, therefore, has to consider the full impact parameter range.
\item
The diffuseness of the neutron periphery and possible differences between two isotopes are neglected.
\item
Usually incident antiprotons do not traverse the skin radially and the absorption probability depends on the impact parameter.
\item
The absorption of produced antihyperons favor the pair production in peripheral reactions.
\item
Different isotopes may also have different proton distributions.
\end{itemize}

Because of the large absorption cross section for antiprotons, the production ratio for $\Sigma^-\overline{\Lambda }$ and $\Lambda \overline{\Lambda }$ pairs reflects the neutron and proton content of the nuclear periphery and will, therefore, be strongly related to the neutron neutron skin or halo. However, one has to keep in mind that particularly at low beam energies the production ratio of $\Sigma^-\overline{\Lambda }$ and $\Lambda \overline{\Lambda }$
pairs is influenced by different production cross sections. Considering, the  double ratio for two isotopes, this dependence largely cancels.
In the following, we will calculate the production ratio between $\Sigma^-\overline{\Lambda }$ and $\Lambda \overline{\Lambda }$  pairs for two isotopes  (see \autoref{eq:dr}) by evaluating the double ratio of the component areal densities for neutrons and protons within one attenuation length for the two considered isotopes.

\begin{figure}[tb]
  \centering
  \includegraphics[width=1.0\columnwidth]{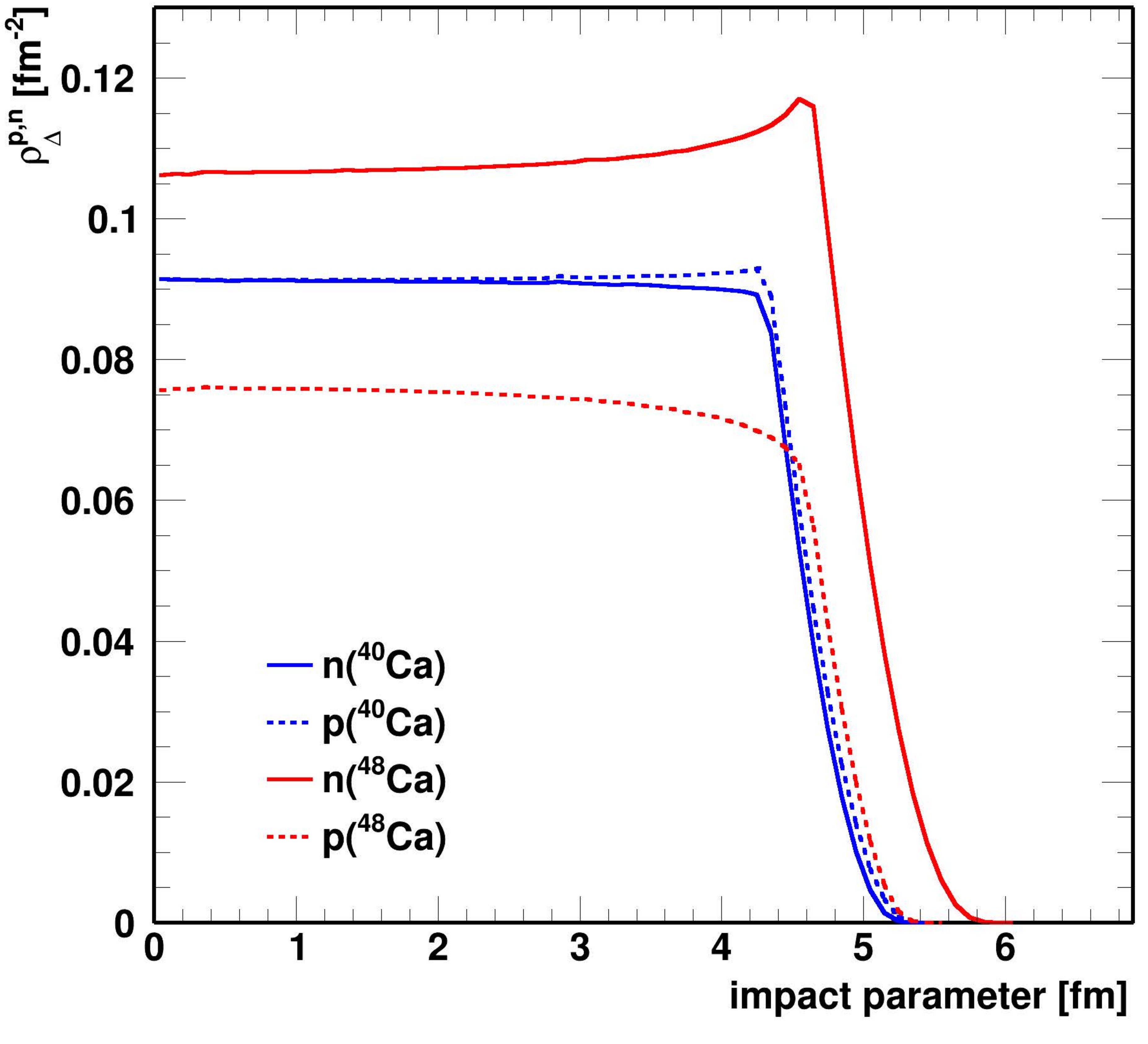} %DensImpact}
  \caption{
  Individual areal densities for protons (dashed lines) and neutrons (solid lines) within one interaction length for
  $^{40}$Ca (blue lines) and  $^{48}$Ca (red lines) at an incident momentum of 2.4\gevc1 according to \autoref{sec_02}.
  The density distributions were generated during the initialization of the Gie\ss en Boltzmann--Uehling--Uhlenbeck (GiBUU) transport model simulations \cite{Buss20121} using the RMF parameter set of Ref.~\cite{PhysRevC.55.540}.
  }
  \label{fig:skinnp}
\end{figure}

In this work we consider an incident $\overline{\text{p}}$ momentum of 2.4 {\gevc1}. At this energy, the interaction cross sections of $\overline{\text{p}}$+p and $\overline{\text{p}}$+n are similar in magnitude and about 55\,mb \cite{Workman:2022ynf}. For such a cross section, the interaction length at normal nuclear density of 0.16\,fm$^{-3}$ is about $\Delta$=1.14\,fm, corresponding to an integrated areal density along the antiproton path of $\int_{\Delta} \rho_\text{n+p}\, \text{d}z \approx $0.18\,fm$^{-2}$. In the periphery, where the density is low, this integral is of course limited by the lower total areal density.

The density distributions used in the following are generated during the initialization of the GiBUU simulations using the RMF parameter set of Ref.~\cite{PhysRevC.55.540}. Assuming for simplicity a straight line trajectory along the z-direction for the incident antiproton, \autoref{fig:skinnp} shows the areal density $\rho_{\Delta}^{q}(b)= \int_{\Delta} \rho_\text{q} dz$ for protons (q=p; red lines) and neutrons (q=n; blue lines) along the antiproton path within one interaction length $\Delta$ for $^{40}$Ca (dashed lines) and  $^{48}$Ca (solid lines) as a function of the impact parameter b.
Of course, at large impact parameters, where the nuclear density is low, a full interaction length can not be reached and the integrated areal densities are lower.
Whereas for  $^{40}$Ca the proton and neutron content within $\Delta$ is rather similar, for $^{48}$Ca the neutrons exceed the protons by more than 50\%.
Integrating the individual components for protons (q=p) or neutrons (q=n) over all impact parameters
\begin{equation}
P^{q}=\frac{\int_{b=0}^{\infty}\rho_{\Delta}^{q}(b){\cdot}b{\cdot}db}{\int_{b=0}^{\infty} b{\cdot}db}
\end{equation}
we find a neutron-to-proton ratio $P^n/P^p$ = 0.97 and 1.65 for $^{40}$Ca and $^{48}$Ca, respectively.
The expected double ratio amounts then to 1.65/0.97 $\approx$ 1.70. The last column in \autoref{tab:skin} lists the double ratios expected for all studied isotope pairs and different RMF parameters from this schematic scenario.

\begin{figure}[tb]
  \centering
  \includegraphics[width=1.0\columnwidth]{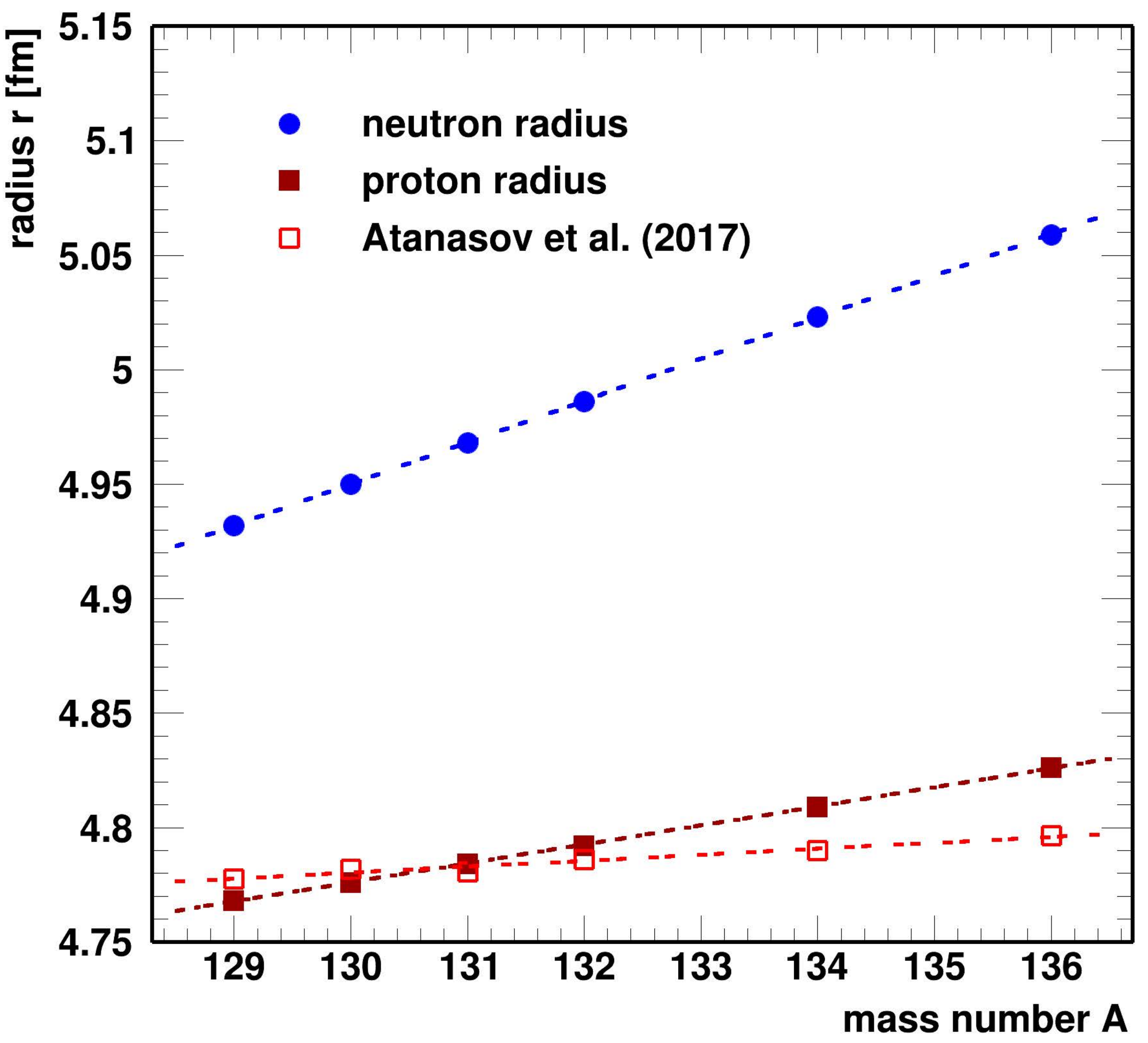} %neutronSkin}
  \caption{Proton (filled red squares) and neutron (blue points) rms radii of the initial distributions for stable or long-lived xenon isotopes used by the GiBUU simulations \cite{Buss20121}. The open red squares show experimental values for the proton radii~\cite{ANGELI201369,Atanasov_2017}. The lines are drawn to guide the eye.
  }
  \label{fig:BUUradii}
\end{figure}
To illustrate the sensitivity of the method in case of only small variations of the neutron distribution, we explore here first the isotope chain of xenon.
At an antiproton storage ring, xenon can be experimentally explored in the mass range from $A=129$ to $A=136$ . Of course, one has to keep in mind, that an efficient gas recirculation system will be mandatory. The filled red squares and blue points in \autoref{fig:BUUradii} show the rms radii of the proton and neutron distributions as a function of the Xe mass number. In these calculations not only the neutron radii but, unlike in the simple geometrical picture presented above also the proton radii are rising slightly with increasing neutron number. This increasing charge radius is qualitatively consistent with experimental data, see open squares in \autoref{fig:BUUradii} \cite{ANGELI201369}, though the experimental slope with the mass number is only about half as large. Note that the range of the neutron skin varies by only 0.13\,fm, which is only a factor of two larger as the uncertainty of the PREX measurement for Pb. Since parity violating e$^-$ scattering is generally hampered by limited statistics, exploring such small neutron skin variation in heavy isotope chains clearly calls for alternative methods.

\begin{figure}[tb]
  \centering
  \includegraphics[width=1.0\columnwidth]{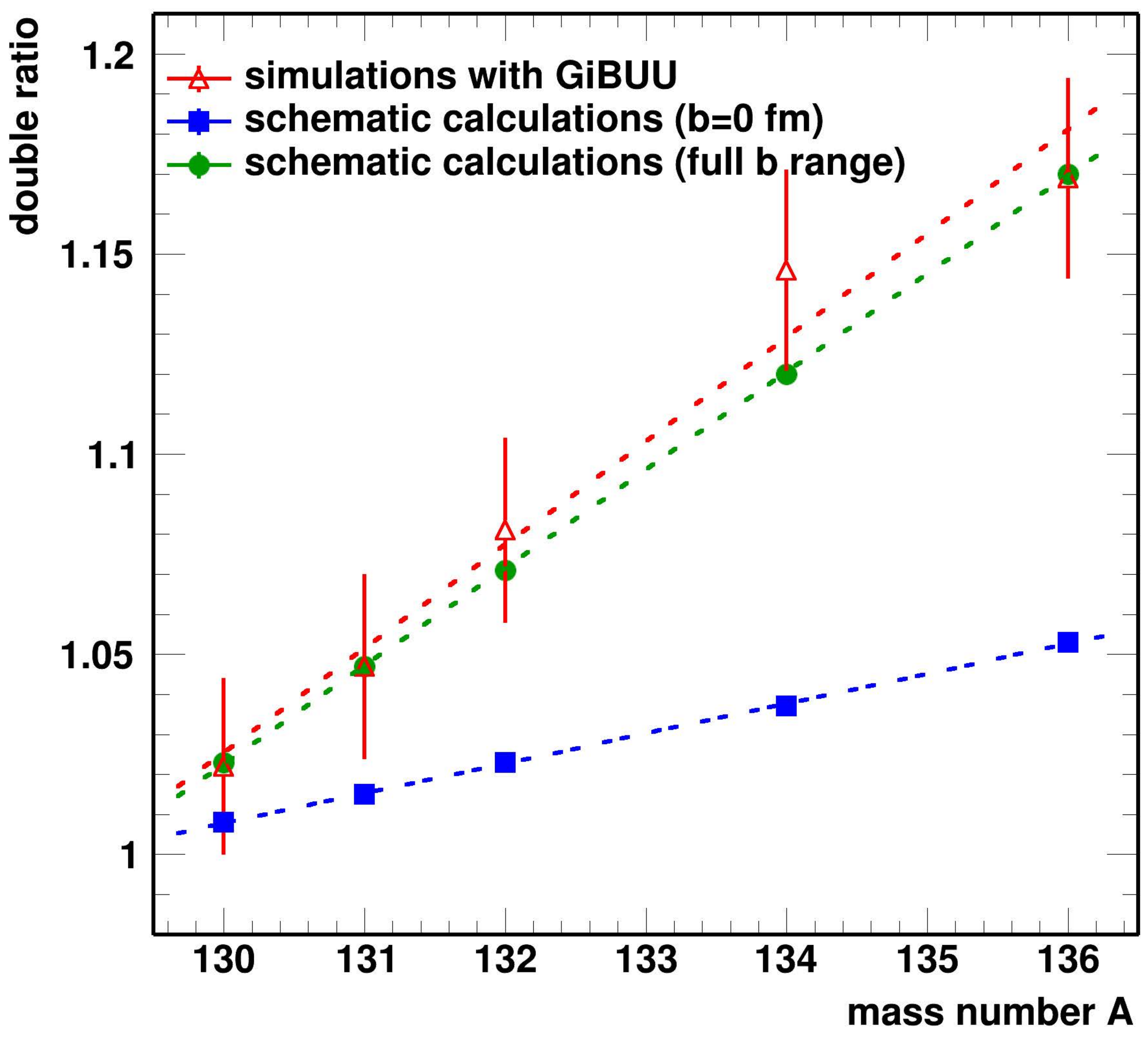} %doubleRatio_3}
  \caption{Double ratio predicted from the two simplified scenarios in \autoref{sec_01} (blue squares) and \autoref{sec_02} (green dots)
  for different xenon isotopes of mass A=130-136 with respect to $^{129}$Xe. The density distributions were generated during the initialization of the Gie\ss en Boltzmann--Uehling--Uhlenbeck (GiBUU) transport model simulations \cite{Buss20121} using the RMF parameter set of Ref.~\cite{PhysRevC.55.540}. The red triangle show the results of the GiBUU simulations.
  The lines are drawn to guide the eye.
  }
  \label{fig:doubleXe}
\end{figure}

\autoref{fig:doubleXe} shows the double ratio from the two simplified scenarios in \autoref{sec_01} (blue squares) and \autoref{sec_02} (green dots) for xenon isotopes with mass A=130-136 with respect to $^{129}$Xe. Compared to the simple analytic expression of \autoref{eq:dr01} with zero impact parameter, the ratio is significantly enhanced by considering the full impact parameter range. This is caused by the larger role played by the neutron rich nuclear periphery at large impact parameters. Within this picture a significant variation of the double ratio over such an isotope chain is expected.

%--------------------------------------
\begin{table*}[htbp]
\begin{ruledtabular}
\centering
\begin{tabular}{c c r r r r r r}

{\bf target}   & {\bf RMF} & {\bf events} & \multicolumn{2}{c}{\bf number of pairs} & \multicolumn{3}{c}{\bf double ratio}\\
               & {\bf model} & {\bf [10$^6$]}     &    {\bf \PgL\PagL}            &     {\bf \PgSm\PagL}             & {\bf GiBUU}&\bf{\bf \autoref{eq:dr01}}& {\bf $\bf{\rho_{\Delta}}$} \bf(\bf\autoref{sec_02})\\

\hline
$^{20}$Ne &  NL3 \cite{PhysRevC.55.540}& 167  & 32387 & 10870 & --- & ---& ---\\
$^{22}$Ne &  NL3 \cite{PhysRevC.55.540}& 171  & 29227 & 13297 & 1.356$\pm$0.021& 1.100& 1.291\\

\hline
$^{40}$Ca &NL1 \cite{PhysRevC.55.540}   &  415  & 76323 &  22880& --- & ---& ---\\
$^{48}$Ca &NL1 \cite{PhysRevC.55.540}   &  450  & 66694 &  36074& 1.799$\pm$0.018& 1.225& 1.683\\

$^{40}$Ca &NL3 \cite{PhysRevC.55.540}   &  415  & 74280 &  21827& ---& ---& ---\\
$^{48}$Ca &NL3 \cite{PhysRevC.55.540}   &  450  & 61313 &  32391& 1.798$\pm$0.019& 1.207& 1.683\\

$^{40}$Ca &NL3* \cite{LALAZISSIS200936} &  415  & 78753 &  23438& ---& ---& ---\\
$^{48}$Ca &NL3* \cite{LALAZISSIS200936} &  450  & 64212 &  34523& 1.807$\pm$0.018& 1.210& 1.685\\

%$^{40}$Ca &Set I \cite{PhysRevC.65.045201} & 415  & 67000 &  17446& ---& ---& ---\\
%$^{48}$Ca &Set I \cite{PhysRevC.65.045201} & 450  & 57016 &  26042& 1.754$\pm$0.020& 1.142& 1.663\\
\hline

$^{58}$Ni &NL3 \cite{PhysRevC.55.540}     &   100  & 16811 &  5230& ---& ---& ---\\
$^{64}$Ni &NL3 \cite{PhysRevC.55.540}     &   108  & 14978 &  6534& 1.402$\pm$0.030& 1.109& 1.340\\

\hline
$^{129}$Xe &  NL3 \cite{PhysRevC.55.540}   & 109  & 13717 &  6238& ---& ---& ---\\
$^{130}$Xe &  NL3 \cite{PhysRevC.55.540}   & 109  & 13394 &  6225& 1.022$\pm$0.022& 1.008& 1.023\\
$^{131}$Xe &  NL3 \cite{PhysRevC.55.540}   & 109  & 13403 &  6379& 1.047$\pm$0.023& 1.015& 1.047\\
$^{132}$Xe &  NL3 \cite{PhysRevC.55.540}   & 109  & 13335 &  6556& 1.081$\pm$0.023& 1.023& 1.071\\
$^{134}$Xe &  NL3 \cite{PhysRevC.55.540}   & 109  & 12771 &  6656& 1.146$\pm$0.025& 1.037& 1.120\\
$^{136}$Xe &  NL3 \cite{PhysRevC.55.540}   & 109  & 12680 &  6739& 1.169$\pm$0.025& 1.053& 1.170\\
\end{tabular}
 \caption{Number of generated inclusive interactions and production yield of $\PgL\PagL$ and $\PgSm\PagL$ pairs in $\Pap$--Ne, Ca, Ni and Xe interactions at an incident momentum of 2.4 {\gevc1}. The last three columns show the double ratios deduced from the GiBUU simulation \cite{Buss20121}, and the simple models presented in \autoref{sec_01} and \autoref{sec_02}. \label{tab:skin}
 }
 \end{ruledtabular}
\end{table*}
%------------------------------------------------

\section{GiBUU transport study \label{sec_03}}

A more realistic description of the hyperon pair production can be achieved by microscopic transport calculations. One such model, the Giessen Boltzmann-Uehling-Uhlenbeck (GiBUU) transport model \cite{Buss20121}, describes many features of \Pagp--nucleus interactions in the FAIR energy range \cite{PhysRevC.80.021601,Buss20121,PhysRevC.85.024614}. Particularly the presently available data on strangeness production are well reproduced. These simulations also avoid the approximations adopted in the derivation of \autoref{eq:dr01}.
In future studies we will also consider the co-planarity of the produced hyperon-antihyperon pair to extract additional information. However, in the present work we restrict the discussion to the total production ratios.
%Preliminary results for neon has been presented earlier \cite{doi:10.7566/JPSCP.17.091002}.

In the following, we employ the GiBUU model to study $\overline{\text{p}}+\text{Ne, Ca, Ni, and Xe}$ reactions of several isotopes.
All studied isotopes have a sizable abundance (\autoref{tab:radii}) and can in principle be used for experimental studies. The simulations were performed with an incident antiproton momentum of 2.4 {\gevc1}.

The number of generated events for each nucleus are listed in \autoref{tab:skin}. Typically, 100 million events require a computational time of about four days on the MOGON2 high performance computing cluster at the University of Mainz. During the initialization of these simulations, the proton and neutron distributions of the target nuclei are generated by a self consistent relativistic mean-field (RMF) model. If not mentioned otherwise, we use the RMF parameter set of Ref.~\cite{PhysRevC.55.540}.
%----------------------------------------------------------------------------

\subsection{The case of xenon isotope chain \label{sec:xenon}}

\begin{figure}[tb]
\includegraphics[width=1.0\columnwidth]{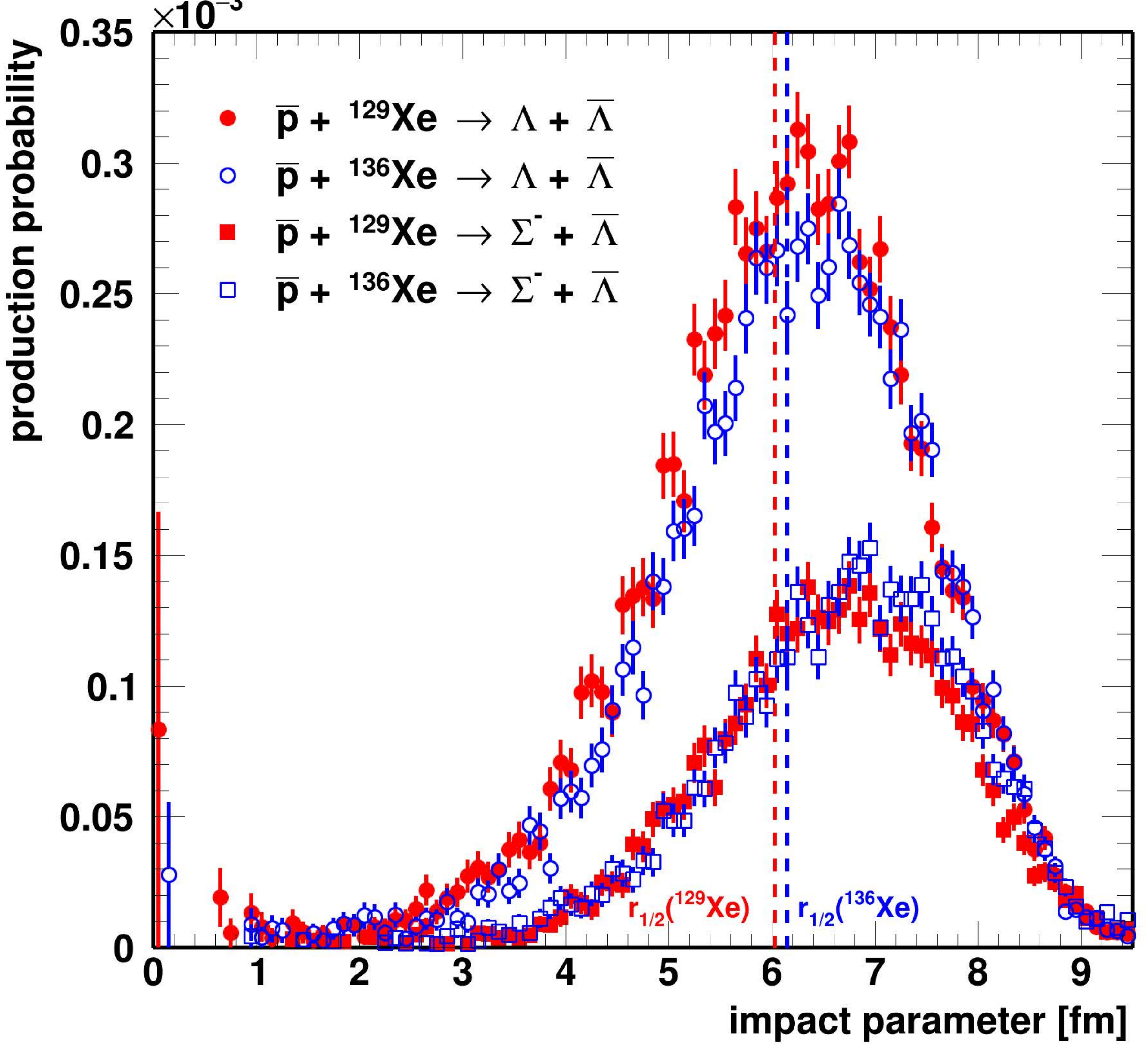}
\caption{
Production probability of $\PgL\PagL$ pairs (circles) and $\PgSm\PagL$ pairs (squares) predicted by the GiBUU transport model \cite{Buss20121}
in exclusive 2.4 GeV/{\em{c}} $\Pap$-$^{129}$Xe (red symbols) and  $\Pap$-$^{136}$Xe (blue symbols) interactions as a function of the impact parameter. For orientation, the two lines mark the half-density radii r$_{1/2}$ of $^{129}$Xe (red line) and $^{136}$Xe (blue line).}
\label{fig:prob}
\end{figure}

\autoref{fig:prob} shows the predicted production probability $\sim$b$^{-1}dN_{\Py\PagL}/db$ for $\PgL\PagL$ pairs (circles) and $\PgSm\PagL$ pairs (squares) as a function of the impact parameter b for for $^{129}$Xe (red) and $^{136}$Xe (blue) interactions.
The vertical dashed lines indicates rms-radii for neutrons and protons of $^{129}$Xe (red) and$^{136}$Xe (blue; see \autoref{tab:radii}).
As expected, the pair production probability is largest in the nuclear periphery, where the chance for both, the $\PagL$  and the $\PgL$ or $\PgSm$ to escape, is sizable. At very small impact parameters particularly the forward going $\PagL$ have to cross a large part of the target nucleus to be emitted. Consequently, for heavy nuclei like xenon the absorption probability for $\PagL$ approaches 1 for more central collisions.

\begin{figure}[b]
\includegraphics[width=1.0\columnwidth]{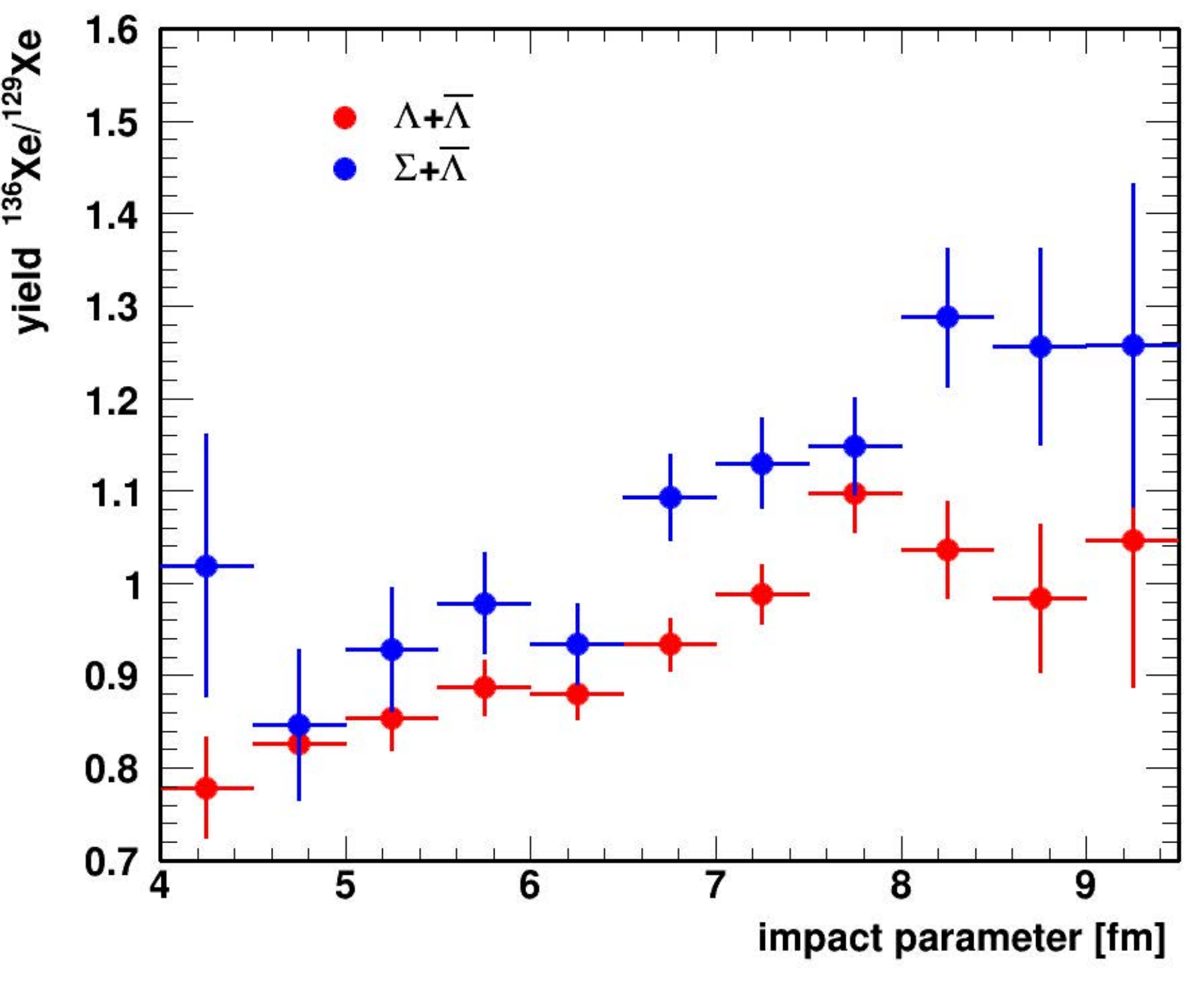}
\caption{Ratio of exclusive $\PgL\PagL$ (red) and $\PgSm\PagL$ pair (blue) production in $^{136}$Xe vs. $^{129}$Xe nuclei as a function of the impact parameter.}
\label{fig:skin}
\end{figure}

\autoref{fig:skin} shows the ratio of exclusive $\PgL\PagL$ (red symbols) and $\PgSm\PagL$ pair (blue symbols) production in $^{136}$Xe vs. $^{129}$Xe nuclei as a function of the impact parameter. At impact parameters around the nuclear radius, the $\PgL\PagL$ production in $\overline{\text{p}} +\text{p}$ interactions is reduced for the more neutron rich isotope due to the more extended neutron skin which leads to an enhanced absorption for the incoming $\overline{\text{p}}$ as well as the outgoing $\PagL$.
At very large impact parameters beyond $>$\SI{7}{\femto\m}, and at hence low matter density, the antiprotons are hardly absorbed and the antiprotons pass the full proton (and neutron) distributions. Since the proton distributions are quite similar for the two isotopes, the $\PgL\PagL$ ratio approaches 1.

The production ratio for $\PgSm\PagL$ pairs indicated by blue symbols in \autoref{fig:skin} shows a somewhat different impact parameter dependence.
For $\PgSm\PagL$ production, the absorptions of the incident antiprotons and the produced $\PagL$ within an extended neutron skin act in opposite direction.
Indeed, a weak suppression is observed at more central collisions and the region of intermediate impact parameters shows similar yields for $^{129}$Xe and $^{136}$Xe.
However, going further into the low density periphery of the nuclei beyond b $>$\SI{6}{\femto\m}, the absorption of the $\PagL$ is less important
and the additional neutron content of $^{136}$Xe enhances the $\PgSm\PagL$ production with respect to $^{129}$Xe considerably.

%---------------------------------------------------------------------

\autoref{tab:skin} gives the individual yields for all six isotopes and the yield ratios between $^{130...136}$Xe and $^{129}$Xe.
Already the individual yields show a continuous sensitivity to the additional neutron layer in $^{136}$Xe.
Since the yield ratios R for $\PgL\PagL$ and $\PgSm\PagL$ pairs are affected in opposite directions, the sensitivity is even more enhanced by forming the double ratio DR of \autoref{eq:dr}. Despite the small neutron skin variations between $^{129}$Xe and $^{136}$Xe by only 0.07\,fm (see \autoref{tab:radii}), the skin variation correlates strongly with the double ratio calculated with the GiBUU code with a Pearson correlation of 0.984 for these Xenon data.

It is interesting to note, that the double ratio from the GiBUU simulations (red triangles in \autoref{fig:doubleXe}) and the ratios from our simplified calculations of \autoref{sec_02} (green dots in \autoref{fig:doubleXe}) agree within errors. Although
the quantitative agreement may be fortuitous, this strong correlation shows that the simple analytic model outlined in \autoref{sec_02} covers the main features of the process. This will be discussed in more detail at the end of this section and in \autoref{sec:sensitivity}.
%----------------------------------------------------------------------------

\subsection{The case of $^{20}$Ne and $^{22}$Ne \label{sec:neon}}

Like the noble gas xenon, also neon is a conceivable element for the \panda cluster-jet target. Besides the common $^{20}$Ne, also
$^{22}$Ne might be a feasible target if the cluster target system is equipped with a gas regeneration system. The measured charge radii of $^{20}$Ne (r$_c$=2.992$\pm$0.008\,fm) and $^{22}$Ne (r$_c$=2.986$\pm$0.021\,fm) differ only very little \cite{DEVRIES1987495}. This small variation is also reproduced by Hatree-Fock-Bogoliubov \cite{GRUMMER1996673} and relativistic mean field calculations \cite{PandaSharma}. On the other hand according to these calculations, the radius of the neutron distribution of $^{22}$Ne is about 0.2\,fm wider than the one of $^{20}$Ne \cite{GRUMMER1996673}. The RMF model implemented in the GiBUU code gives a difference which is about half as large (see \autoref{tab:radii})

The upper part of \autoref{tab:skin} gives the individual yields and the yield ratios for $^{22}$Ne and $^{20}$Ne targets predicted by GiBUU for an incident antiproton momentum of 2.4\gevc1. Unlike for the xenon isotopes, already the individual yields show a remarkable sensitivity to the additional neutron layer in $^{22}$Ne. Since the yield ratios for $\PgL\PagL$ and $\PgSm\PagL$-pairs are affected in opposite directions by the additional neutron layer, the sensitivity is even more enhanced by forming the double ratio DR, which reaches a value around 1.4.

%---------------------------------------------------------------------
\subsection{The case of $^{40}$Ca and $^{48}$Ca \label{sec:calcium}}

\begin{figure}[bp]
  \centering
  \includegraphics[width=1.0\columnwidth]{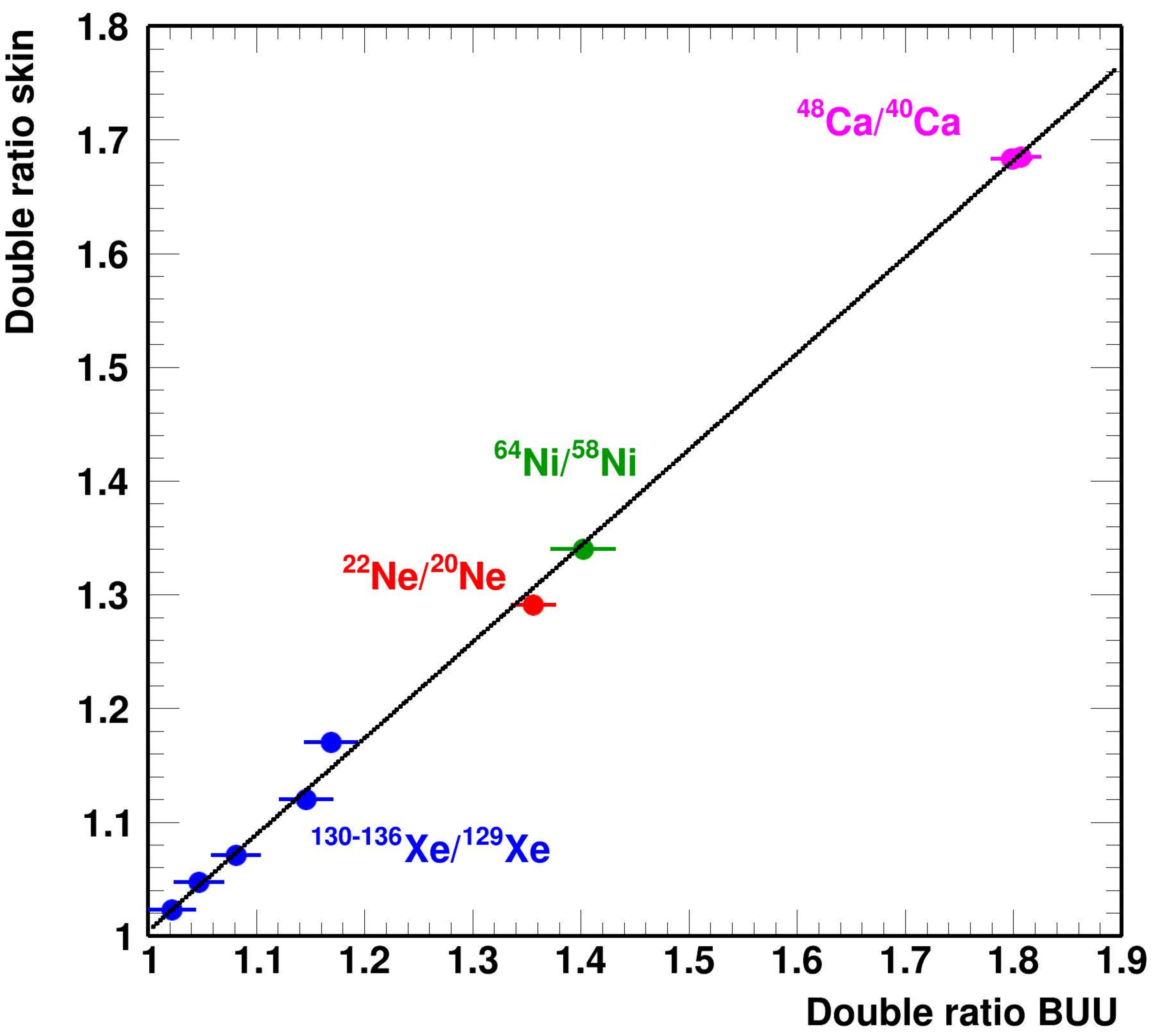} %pbar_ratio}
  \caption{
  Double ratio deduced from the neutron-proton content of the nuclear periphery as described in \autoref{sec_02} as a function of the
  double ratio predicted by the GiBUU transport calculations of 2.4{\gevc1}  $\overline{\text{p}}$+A interactions. The black line is a linear fit to the data. The Pearson coefficient between the two double rations amounts to 0.999.
  }
  \label{fig:dr_buuskin}
\end{figure}

Charge radii of calcium isotopes are very well known \cite{Miller2019}. Indeed, $^{48}$Ca has the same rms charge radius as $^{40}$Ca within 0.001$\pm$0.003(stat.)$\pm$0.010(syst.)fm \cite{GarciaRuiz2016}, making this isotope pair
%at first sight
an ideal case for the proposed method.
One has to keep in mind though, that the proton density {\em distributions} also show significant differences \cite{EMRICH1983401,Hodgson1992}.

The GiBUU model evaluates the initial charge and neutron distributions of the target nucleus with a specified RMF parameter set. The same parametrization also enters during the time evolution of the reaction. Of course, any interpretation of experimental data should be constrained to simulations, which describe the proton distributions (and other possible observables) reasonably well.
In the present work we use three RMF descriptions, the NL1 \cite{PhysRevC.55.540}, NL3, and NL3* \cite{LALAZISSIS200936}, which provide slightly different proton and neutron distributions. In all three cases the proton radii are close to the experimental values(see \autoref{tab:radii}).
In line with the similar proton and neutron distributions, the double ratios from the GiBUU simulations agree within statistical uncertainties for all three cases (see \autoref{tab:skin}).

In \autoref{fig:dr_buuskin} we depict the double ratio deduced from the neutron-proton content of the nuclear periphery as described in \autoref{sec_02} as a function of the double ratio predicted by the GiBUU transport calculations of 2.4{\gevc1} $\overline{\text{p}}$+A interactions.
The systematic difference between the two double ratios can be at least in part traced back to the differences in absorption of antihyperons and hyperons, which is particularly large at small impact parameters, where the double ratio is smaller (cf. \autoref{fig:skin} and \autoref{fig:skinnp}).  In our schematic model this absorption effect has been neglected.
Nonetheless, the strong correlation between the two quantities in \autoref{fig:dr_buuskin} with a slope close to 0.9 and with a remarkable Pearson coefficient of 0.999 suggests that our schematic model accounts for the main features of the reaction process modeled by the GiBUU code.
The fact, that this linear correlation holds over a wide mass and element range, is also in line with \autoref{eq:dr01b} which itself indicates within the (oversimplified) scenario of \autoref{sec_01} the dominance of the neutron skin difference. It seems, that the double ration is not strongly influenced by the dynamics of the reaction process.

In cases where predictions for the density distributions of protons and neutrons exist, the prescription of \autoref{sec_02} and the correlation of \autoref{fig:dr_buuskin} allow the evaluation of the double ratio which could be confronted with experiments.

\section{Sensitivity and Systematics of the Method\label{sec:sensitivity}}
\subsection{General Aspects}
The results for the xenon isotopes give an important hint on the sensitivity of double ratios to the thickness of the neutron
skin. It is, however, highly desirable to investigate that relation systematically
on more general grounds. Under theoretical aspects, one may use results obtained
by a representative selection of nuclear models. That was the approach pursued in the previous sections: The RMF models incorporated into the GiBUU numerical transport package are well tested and widely and successfully used mean-field models. The results are typical at least for covariant non-linear energy density functionals. Since modern mean-field approaches lead within error bars to the same results for nuclear masses and form factors, the GiBUU-based result discussed in the previous sections can be considered of being valid and realistic even beyond the special class of built-in mean-field approaches. Obviously, extending the list to other covariant or non-covariant energy density functional approaches will not lead to fundamentally different conclusions.

However, it is still of value to dig a bit deeper into the relation between antiproton-nucleus double ratios and the nuclear skin properties. It is worthwhile to recall that
the rms-radius, together with the diffusivity, and the half-density radius, as defining key element of nuclear density distributions, are of high research interest. Most spectacular, this is reflected by unexpected results of the PREX/CREX experiments on the neutron skin in $^{208}$Pb. They are in tension with the seemingly well established former, largely theoretically based knowledge on that issue.

In this paper and especially in this section we anticipate that antiproton annihilation on nuclei provides an independent approach on nuclear skin research. For that aim, we let aside for the moment the numerically involved self-consisted RMF description and revert to a easy to handle, but realistic, schematic approach. Except for the lightest nuclei, nuclear density distributions are well described by form factors of Wood-Saxon shape. Proton ($A_q=Z$) and neutron ($A_q=N\equiv A-Z$) number density distributions are parameterized by Fermi-distributions
\begin{equation}
\label{eq:fermi}
\rho_{q}(r)=\frac{\rho_0(A_q)}{1+e^{(r-R_q)/a_q}}
\end{equation}
where for simplicity we assume spherical symmetry.
The half-density radius $R_q$ and diffuseness $a_q$ may be adjusted to electron scattering data or to theoretical RMF or non-relativistic HFB results, respectively. Normalization to the respective mass and charge numbers leads to
\begin{equation}
\label{eq:rho0WS}
\rho_0(A_q)=\frac{3A_q}{4\pi R^3_q(1+x_q)},
\end{equation}
where $x_q=(\frac{\pi a_q}{R_q})^2$  and $\rho_0(A_q)\sim \rho_{q}(r)_{|r=0}$ is the density at the center of the nucleus.
The mean-square radius is
\begin{equation}
\label{eq:rmsWS}
\langle r^2_q\rangle = \frac{3 }{5}R^2_q\left(\frac{1+\frac{121}{30} x_q +\frac{7}{3} x^{2}_q}{ 1+x_q}\right).
\end{equation}
The expressions of Eq.\eqref{eq:rho0WS} and Eq.\eqref{eq:rmsWS} are correct up to terms of order $\mathcal{O}(e^{-\frac{R_q}{a_q}})$. For $a_q=0$ the expressions of Eq.\eqref{eq:rho0WS} and Eq.\eqref{eq:rho0WS} reduce to the respective hard sphere results which we adopted in \autoref{sec_01}.
The two central messages of this exercise are firstly, that the nuclear geometrical parameters are in fact intimately and non-linearly correlated and, secondly, that strict constraints on parameter variations are imposed by the requirement that the normalization to proton and neutron numbers must be maintained separately. A clear advantage of the Fermi-function model is to illustrate those aspects of nuclear density distributions in a very transparent manner.

In this section we concentrate on the isotope pair of $^{40}$Ca and $^{48}$Ca. To mimic the effect of different nuclear models, we modify the neutron distribution of $^{48}$Ca. In the following, two nominally different scaling approaches will be used to study possible statistical and systematic uncertainties, respectively.

\subsection{Radial Scaling as a Form-Invariant Transformation\label{sec:statistics}}

The Fermi-function model is an ideal tool to clarify the connections between radial scaling and scaling of density parameters.
Radial scaling by a factor $\alpha$ amounts to stretch ($\alpha > 1$) or squeeze ($0< \alpha < 1$) the radial coordinate:
\begin{equation}
\rho_q(r)\to \widetilde{\rho}_q(\alpha r)
\end{equation}
constraint by
\begin{equation}
A_q=\int d^3r \rho_q(r)=\int d^3r_\alpha \widetilde{\rho}_q(r_\alpha)
\end{equation}
which corresponds to  require equality of the integrands, $d^3r \rho_q(r)=d^3r_\alpha \widetilde{\rho}_q(r_\alpha)$.
The norm integral is conserved by using $\widetilde{\rho}_q(\alpha r)=\rho_q(\alpha r)$ and $d^3r_\alpha = \alpha^3 d^3r$, or
alternatively, $\widetilde{\rho}_q(r_\alpha)=\alpha^3\rho_q(\alpha r)$ combined with the unchanged volume element
$d^3r_\alpha = d^3r$.

Modelling the nucleon density distributions by Fermi-functions, radial scaling produces
\begin{equation}
\rho_q(\alpha r)=)
\frac{\rho_0(A_q)}{1+e^{(\alpha r-R_q)/a_q}}=\frac{\rho_0(A_q)}{1+e^{(r-R^{(\alpha)}_q)/a^{(\alpha)}_q}}
\end{equation}
with $R^{(\alpha)}_q=R_q/\alpha$ and $a^{(\alpha)}_q=a_q/\alpha$. Hence, radial scaling corresponds to scale simultaneously the half-density radius $R_q$ and the diffusivity $a_q$ by the same factor.
Hence, we are led to the conclusion that radial scaling is a rather restrictive approach because mathematically it is a shape-invariant transformation.
Still, it can be used to estimate statistical uncertainties of the method.

Of course, such a radial scaling can be applied to any distribution. For this study we make use of density distributions which are generated with the initialization code of the GiBUU simulations \cite{Buss20121} using different RMF parameter sets of Ref.~\cite{PhysRevC.55.540}. We consider here only parameter sets which give proton radii in good agreement with the experimental values (see \autoref{tab:radii}).
\begin{figure}[tb]
  \centering
  \includegraphics[width=1.0\columnwidth]{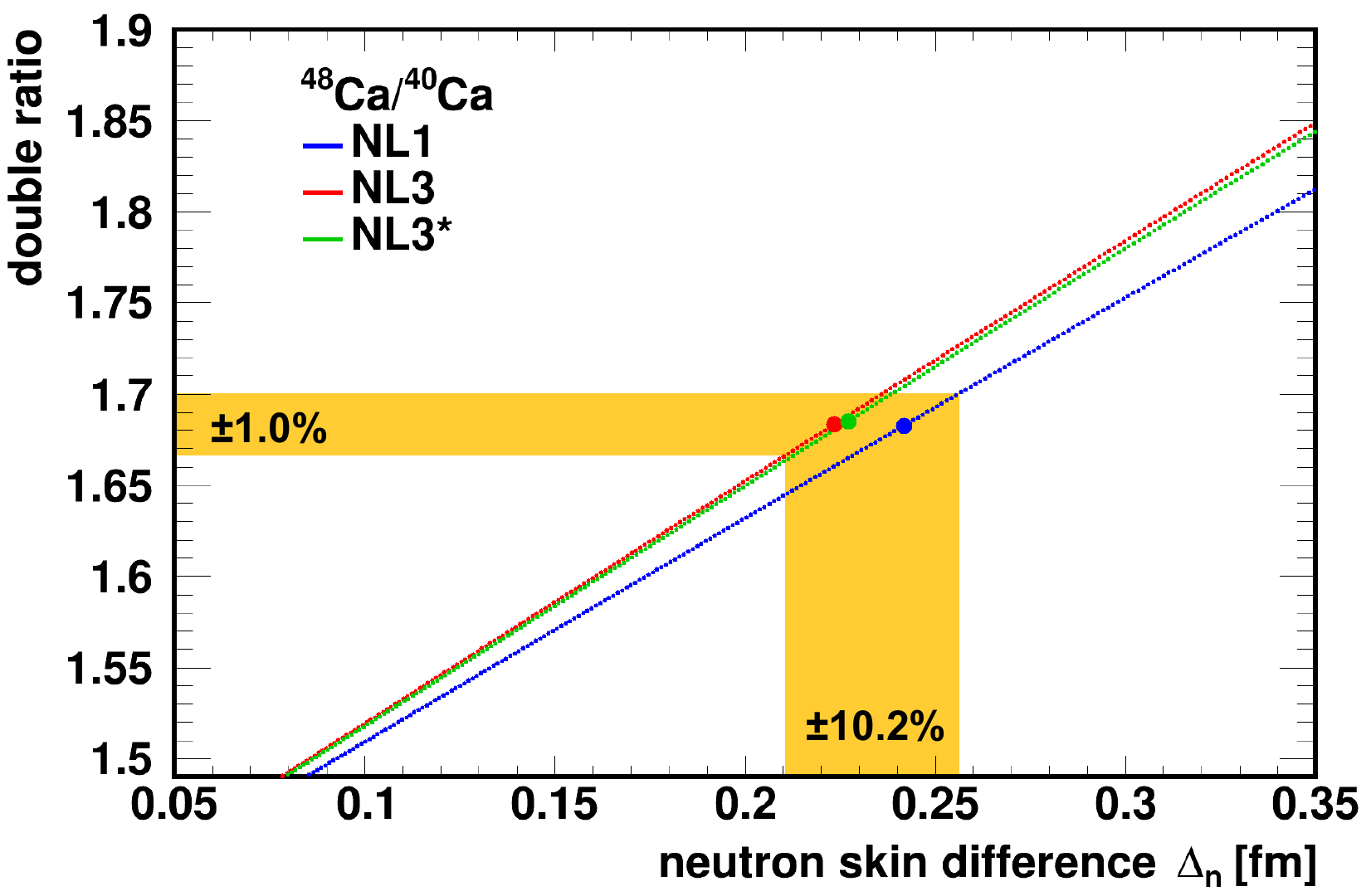} %DifRatio3}    %  sensitivity_02}   %
  \caption{
  Double ratio deduced from the neutron-proton content of the nuclear periphery versus the difference of the neutron skin thickness $\Delta_n$ for $^{40}$Ca and  $^{48}$Ca. In order to explore the effect of different neutron skins for $^{48}$Ca, its neutron
  distribution generated by GiBUU was artificially scaled in radial direction by a given factor, while the proton distribution and the distributions for  $^{40}$Ca remained unchanged. The density distributions are generated during the initialization of the GiBUU simulations \cite{Buss20121} using the RMF parameter set of Ref.~\cite{PhysRevC.55.540}.
  }
  \label{fig:skindr}
\end{figure}
First, the neutron density distribution of $^{48}$Ca was radially scaled. Subsequently, the neutron density distribution was renormalized to the total number of neutrons of 28. As stressed before, the diffuseness of the neutron skin is scaled correspondingly. The neutron distribution of $^{40}$Ca and the proton distributions of both isotopes remained unchanged.

\autoref{fig:skindr} shows the double ratio calculated as described in \autoref{sec_02} versus the difference $\Delta_n$ of the neutron skin difference between the $^{40}$Ca and the modified $^{48}$Ca nucleus, $\Delta_n$ = ${\Delta}R_{pn}$($^{48}$Ca)-${\Delta}R_{pn}$($^{40}$Ca). The symbols on each line mark the double ratio and neutron skin difference for the original density distributions (c.f. \autoref{tab:radii}). For all three interaction parameters we find a very similar linear relation between the double ratio and $\Delta_n$. An uncertainty of the double ratio of $\pm$1\%, which is typical for the GiBUU data sets in this work (see \autoref{tab:skin}) translates into an uncertainty of the neutron skin variation by about $\pm$7\%.
Including possible systematic variations due to the three different interactions, an uncertainty of the double ratio of $\pm$1\% translates into an uncertainty of the neutron skin variation by about $\pm$10\% (see yellow bands in \autoref{fig:skindr}). This is about a factor of three smaller compared to the uncertainty of the CREX result.

\subsection{Two-Parameter Scan in the Fermi-Model\label{sec:woodsaxon}}
If nuclear density distributions would obey radial scaling, then knowing one nuclear density would be sufficient for knowing the density distributions of all nuclei. Obviously, that is not the case. In reality, nuclear density distributions depend on the properties of wave function, seen most impressively in neutron-rich halo nuclei. Thus, half-density radius and diffuseness evolve with proton and neutron number by their own rules, defined by the shell structure, i.e. single particle angular momentum, and separation energies which are determined by the interplay of the various components of the nuclear mean-field, including long-range Coulomb interactions.

%Non-trivial variations of microscopically defined density distributions are only obtained by explicit manipulation of the model parameters, e.g. the coupling constants of the underlying energy density functional. Such readjustments, however, will affect all other observables, although the densities will always be normalized correctly to the respective particle numbers because they are constructed  by the ground state expectation value of the one-body density matrix in terms of single particle wave functions.

The Fermi-function approach provides the degrees of freedom necessary for realistically modelling nuclear density distribution. The general case is given by varying independently $R_q$ and $a_q$:
\begin{equation}
\rho^{(\lambda,\kappa)}_q=\rho^{(\lambda,\kappa)}_0(A_q)\frac{1}{1+e^{(r-R^{(\lambda)}_q)/a^{(\kappa)}_q}}
\end{equation}
where $\rho^{(\lambda,\kappa)}_0(A_q)$ is defined with the scaled parameters $R^{(\lambda)}_q=\lambda R_q$ and
$a^{(\kappa)}_q=\kappa a_q$, assuring the proper normalization. Evidently, the above ansatz allows to study shape variations, induced by changes in $R_q$ and/or $a_q$.

\begin{figure}[tb]
  \centering
  \includegraphics[width=1.0\columnwidth]{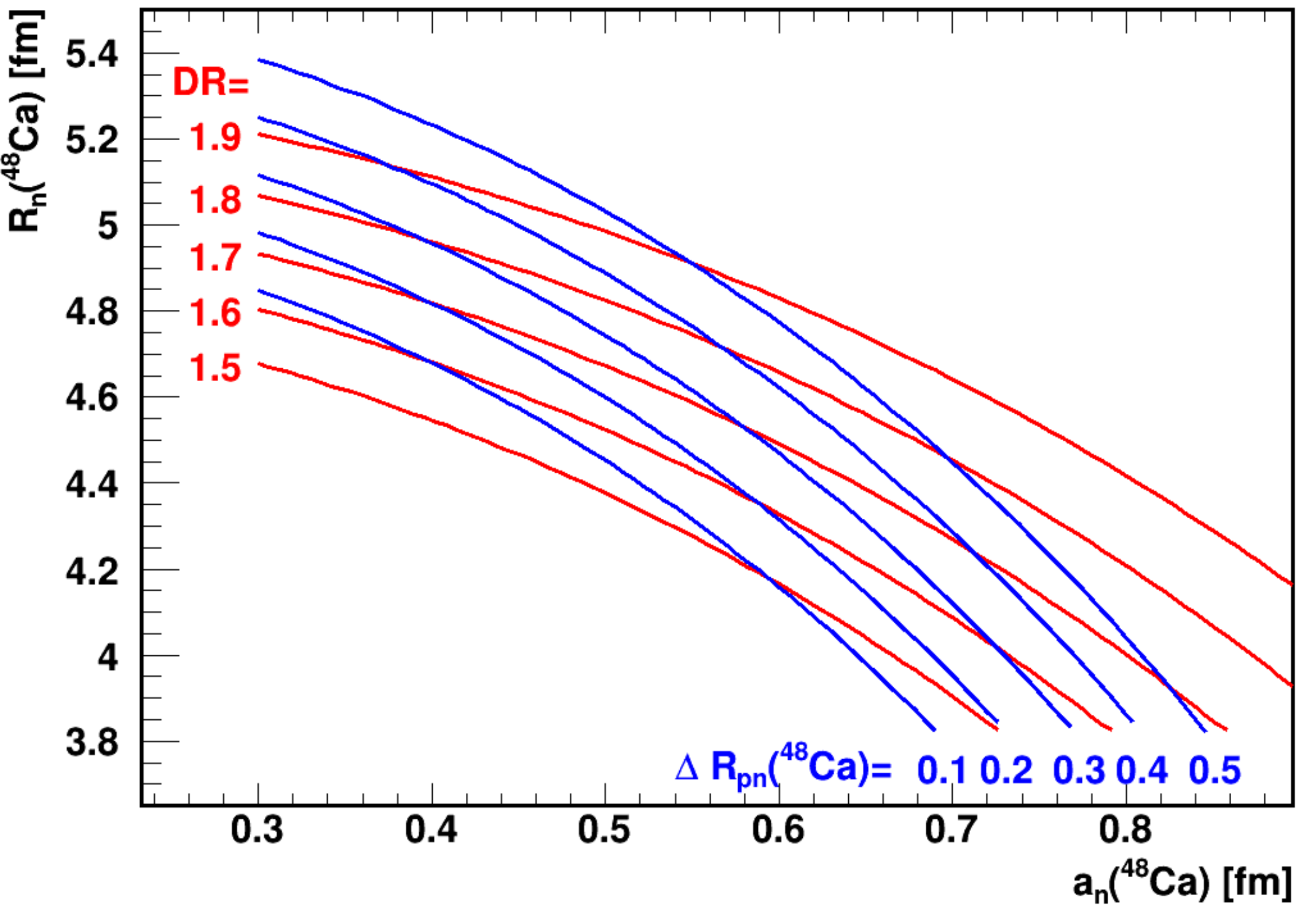}
  \caption{
  Probing the dependence of the Double Ratio DR and the neutron skin thickness $\Delta$R$_{pn}$ on variations of the half-density radius R$_n$ and the diffuseness a$_n$ of the neutron density distribution in $^{48}$Ca. Calculations were performed according to \autoref{sec_02} with a two-parameter
  Fermi function as an anzatz for the density distributions of $^{48}$Ca with half-density radius R$_n$($^{48}$Ca) and diffuseness a$_n$($^{48}$Ca)
  For the proton distributions of $^{40}$Ca and $^{48}$Ca and the neutron distribution of $^{40}$Ca fixed half-density radii of 4\,fm and diffuseness of 0.6\,fm were used. The plot shows isolines for the double ratio (blue) and neutron skin thickness of $^{48}$Ca (red).
  }
  \label{fig:wood}
\end{figure}

As pointed out in our introduction, the characterization of the neutron density distributions in terms of a single quantity like the root-mean-squared radius is a major simplification. In fact, differing nuclear density distributions may lead to the same rms radius.
This can be nicely illustrated by the two-parameter Fermi distribution function, where the rms radius depends on both, the radius and the diffuseness of the distribution (see \autoref{eq:rmsWS}). Requiring for example a fixed R$_{rms}$=4\,fm,  a variation of the diffuseness parameter by $\Delta$a$_q=\pm$0.05\,fm can be compensated by a variation of $\Delta$R$_q\approx\mp$0.12\,fm and vice versa. For orientation, the diffuseness parameters of $^{40}$Ca and $^{48}$Ca are expected in the range from 0.45 to 0.55\,fm (see e.g. \cite{10.1093/ptep/ptab136}).

We will make use of the model of \autoref{sec_02} to illustrate the sensitivity of the double ratio to both parameters of a two-parameter Fermi distribution. For the proton distributions of $^{40}$Ca and $^{48}$Ca and the neutron distribution of $^{40}$Ca we adopted fixed half-density radii of 4\,fm and diffuseness parameters of 0.6\,fm, which describe the GiBUU generated ones reasonably well. For the neutron distribution of $^{48}$Ca the half-density radius R$_n$($^{48}$Ca) and the skin depth a$_n$($^{48}$Ca)  were varied independently.

\autoref{fig:wood} shows isolines for the double ratio (blue) and neutron skin thickness of $^{48}$Ca (red) defined via the difference of the proton and neutron rms radii, in the R$_n$($^{48}$Ca) - a$_n$($^{48}$Ca) parameter space.
A precise determination of the double ratio will determine a specific narrow monotonic relation between R$_n$ and a$_n$ in \autoref{fig:wood}.
Note, that for a Fermi function the diffuseness parameter corresponds to the inverse of the slope at the half-density radius.
Thus, without referring to a specific analytical shape of the density distributions, the slope (or its inverse) at the half-density radius $d\rho/dr|_{r=R_q}$, may provide a more general characterization of the diffuseness of theoretical predictions.
Since e.g. each model shown in \autoref{fig:CaPb} will mark a point in \autoref{fig:wood}, the double ratio may serve as a sensitive test for the various models.

The fact, that the blue and red isolines are not running parallel signals that the double ratio is not uniquely related to the rms-radius of the neutrons
but also on admixtures of higher moments.
Indeed, the combination of the present method with other observables which probe different regions of the neutron skin like e.g. antiprotonic atoms \cite{PhysRevLett.87.082501}, may enable a deeper view beyond the simple rms value on the neutron skin.

%--------------------------------------------------------------------
\section{Experimental aspects and outlook}

The present GiBUU simulations focused on an incident antiproton of 2.4{\gevc1}. Of course, the beam momentum needs to be optimized with respect to e.g. the production yields, experimental efficiency and sensitivity. In particular, the $\Sigma^-\overline{\Lambda}$ pair production will be reduced at lower beam momenta. Preliminary results for the neon case at a lower momentum of 1.7{\gevc1} have been presented earlier \cite{doi:10.7566/JPSCP.17.091002}. Due to the larger $\overline{\text{p}}$+n cross section of 63\,mb at an antiproton momentum 1.7{\gevc1}, the schematic model of \autoref{sec_02} predicts an increase of the double ratio by about 1.4 \% as compared to the case of 2.4{\gevc1}. The GiBUU simulations (with improved statistics with respect to
\cite{doi:10.7566/JPSCP.17.091002}) predict a double ratio of 1.440$\pm$0.054 and hence a somewhat larger increase by 6.2$\pm$4.1\%. However, the statistical uncertainty is still rather large due to the limited computing time. Though this explorative result suggests that the incident antiproton momentum has only a small effect on the double ratio, a systematic study with larger event samples is clearly desirable.
It would be of interest to perform corresponding calculations with other transport model codes
\cite{PhysRevC.101.014608,PhysRevC.105.014623,YONG2024138662,EHEHALT1996449,GEISS1998107,HARTNACK2012119,PhysRevC.101.044905}, also under the aspect of systematic model uncertainties.

The proposed method focuses on the total production ratios of $\PgL\PagL$ and $\PgSm\PagL$ pairs.
The masses of the  $\PgL$ and $\PagL$ hyperons can be easily reconstructed by the charged decay channel. The identification of the $\PgSm\rightarrow \pi^-n$ decay channel, which dominates with 99.8\%, will be more challenging.
At \panda, about 2.7\% of the $\Sigma^-$ tracks can be reconstructed by the micro vertex detector. The $\Sigma^-$ can be identified by a kinematic fit
including the $\Sigma^--\pi^-$ decay vertex and the associated hit of the neutron in the calorimeter.
The reduced \PgSm\ mass resolution will of course increase the background for $\PgSm\PagL$ pairs.
Nevertheless, in the double ratio, many systematic effects will cancel. Furthermore, $\Pap$--p and $\Pap$--d interactions will provide important points of reference.

At \panda a reconstruction efficiencies of 30\%, 30\%, and 2\% are expected for $\PgL$, $\PagL$, and $\PgSm$, respectively \cite{phdthesisSchupp}. Even with a moderate average interaction rate of 2$\cdot$10$^6$s$^{-1}$, the statistical precision shown in \autoref{tab:skin} e.g. for calcium can be reached in about half a day running for each isotope. Although the additional background particularly for the $\PgSm$ may require somewhat larger measuring periods, this estimate demonstrates the large potential of this method for a precision study at an antiproton storage ring.

With the regular pellet target system of the {\panda} setup, all noble gases can be employed. But even with a cleaning and recirculation system the operation of such isotopically enriched gaseous isotopes will be rather costly. Instead, with a filament target system similar to the one, which has been developed for the hyperatom studies at \panda \cite{Steinen2020PhDthesis,SCHUPP2023168684}, also many solid nuclear targets are also feasible. Note, that only a few milligrams of material is needed for such a target filaments. While we have focused in this work on the two double magic calcium isotopes, also additional calcium isotopes or many other isotope chains like e.g. %$^{46-50}$Ti \cite{Yadav_2022} or
$^{58-64}$Ni \cite{PhysRevC.104.034303} will therefore be experimentally accessible at a reasonably price, enrichment \cite{IsotopePrices} and measuring time. For specific Isotope pairs, where sufficient target material is available, also a measurement in a fixed target mode may be feasible, e.g. making use of the antiproton beam at J-PARC \cite{jparc_suzuki}.

Finally we note, that an analogous method might also allow to explore the evolution of the proton skin thickness in isotone chains. Combined with precise data on the proton distributions this might give access to the neutron distributions in proton rich isotones.

\section*{acknowledgments}
We are grateful to Patrick Achenbach for supporting discussions during the early stage of this work.
This project has received funding from the European Union’s Horizon 2020 research and innovation programme under grant agreement No 824093.
The presented data were collected within the framework of the PhD thesis of Falk Schupp \cite{phdthesisSchupp}
and in part within the bachelor thesis of Martin Christiansen \cite{HYP2022_refId23,Christiansen2022BachelorThesis}
at the Johannes Gutenberg University Mainz.
Calculations for this project were performed on the HPC cluster "HIMSter2" at the Helmholtz Institute Mainz. "HIMster2" is a part of Mogon2 and we thank the HPC department of the Johannes Gutenberg University Mainz \cite{mogon2} for their efforts in maintaining the cluster.

F.S. performed the calculations and the analysis. He was responsible for the visualization of the published work and he also took part in the preparation, the review, and the editing of the manuscript. J.P. conceptualized the work, supervised the work and prepared the original draft. M.B. M.C. and M.S. contributed to the data visualization and data collection, T.G. and H.L. helped in implementing the computer code and validation of the results. They also took part in the review and editing process.
\vspace{4cm}

\appendix*
\section{}
The following tables \autoref{tab:nskinCa} and \autoref{tab:nskinPb} collect the values and references for the data displayed in \autoref{fig:CaPb}.
%-------------------------------------------------------------------------------
\begin{table*}[t]
\begin{ruledtabular}
\begin{tabular}{lllll}
{\bf Method}  & {\bf Ref.} & {\bf R$_p$ [fm]}  &{\bf R$_n$ [fm]} &  {\bf ${\Delta }$R$_{pn}$ [fm]}\\
\hline
%1972 &
10.8-16.3\,MeV p elastic scattering & \cite{LOMBARDI1972103}  & 3.38   &  3.78$\pm$0.09 &  0.39$\pm$ 0.10\\
%1972 &
1044\,MeV p elastic scattering & \cite{ALKHAZOV1976443}       & 3.48   &  3.66          &  0.16$\pm$ 0.023\\
%1977 &
1040\,MeV p elastic scattering & \cite{CHAUMEAUX197733}         & 3.38   &  3.54        &  0.16$\pm$ 0.05\\
%1978 &
1040\,MeV p scattering & \cite{BRISSAUD1979141}& 3.41   &  3.58$\pm$0.04$\pm$ 0.1       & 0.18$\pm$0.04$\pm$0.1\\
%1979 &
800\,MeV $\vec{p}$+$^{48}$Ca& \cite{IGO1979151}            & 3.376$^1$          &3.561$^1$   & 0.18$\pm$0.08\\
%1992 &
500-1040\,MeV p scattering & \cite{PhysRevC.67.054605}&    &  3.436 $\pm$ 0.023       & 0.079$\pm$0.023\\
%2017 &
n+$^{48}$Ca and p+$^{48}$Ca scattering & \cite{PhysRevLett.119.222503}&    &    & 0.249$\pm$0.023\\
%2020 &

p+$^{48}$Ca and $^{48}$Ca+$^{12}$C scattering & \cite{TAGAMI2022105155}&  &  & 0.158$\pm$0.023(exp)$\pm$0.012(mod)\\
%2022 &
p+$^{48}$Ca and $^{48}$Ca+$^{12}$C scattering& \cite{PhysRevC.105.014626}&  & 3.43$\pm$0.16 & \\
%2022 &
\hline
166\,MeV $\alpha$ elastic scattering & \cite{BRISSAUD1972145} & 3.39   &  3.72$\pm$0.12 &  0.33$\pm$ 0.12\\
%1976 &
1370\,MeV $\alpha$ elastic scattering & \cite{ALKHAZOV1977365}& 3.48   &  3.67$\pm$ 0.05&  0.19$\pm$ 0.05\\
%1977 &
104\,MeV $\alpha$ elastic scattering& \cite{PhysRevLett.41.1220}&        &              & 0.17$\pm$ 0.10\\
%1979 &
$^{48}$Ca+$^{12}$C interaction cross section & \cite{PhysRevLett.124.102501}&       &       & 0.197$\pm$ 0.048\\
%2020 &

\hline
$\pi^-$ and $\pi^+$ scattering  & \cite{PhysRevLett.38.1201}    &      &                &  0.20$\pm$ 0.09$^1$\\
%1977 &

\hline
$\pi^-$ and $\pi^+$ scattering & \cite{PhysRevC.29.182,PhysRevC.46.1825}& 3.32$\pm$ 0.03& 3.43$\pm$ 0.03 & 0.11$\pm$0.04\\
%2001 &
analysis of antiprotonic atoms & \cite{PhysRevC.65.014306}&    &        & 0.12$^{+0.04}_{-0.08}$\\
%2003 &
\hline
pionic atoms                       & \cite{FRIEDMAN201246}  & &    & 0.13$\pm$0.06 \\ % Friedman 2012
                                   & \cite{FRIEDMAN201246}  & &    & 0.16$\pm$0.07  \\  % Friedmann

\hline
electric dipole polarizability & \cite{PhysRevLett.118.252501} &           &            & 0.17$\pm$ 0.03\\
%2017 &
\hline
CREX  & \cite{PhysRevLett.129.042501} &  &  & 0.121$\pm$0.026(exp)$\pm$0.024(mod) \\
\hline
\multicolumn{5}{l}{$^{1}$\footnotesize{assuming $R_p$=$R_n$ for $^{40}$Ca.}} \\
\end{tabular}
\caption{Published radii of protons and neutrons in $^{48}$Ca which are shown in \autoref{fig:CaPb}.}
\label{tab:nskinCa}
\end{ruledtabular}
\end{table*}
%------------------------------------------------------------------------------

\begin{table*}[tb]
\begin{ruledtabular}
\begin{tabular}{l l l l}
{\bf Method} & {\bf Ref.} &  {\bf ${\Delta }$R$_{pn}$ [fm]} & {\bf remark} \\
\hline
elastic p and n scattering at 40, 65, 200\,MeV & \cite{PhysRevC.65.044306} & 0.17 & \\
elastic p scattering at 295\,MeV   & \cite{PhysRevC.82.044611}      & 0.211$^{+0.054}_{-0.063}$ & \\                 % Zenihiro 2010
elastic p scattering at 650\,MeV   & \cite{PhysRevC.49.2118,PhysRevC.52.291}        & 0.20$\pm$ 0.04 & \\            % Starudubsky 1994
\hline
p+$^{208}$Pb reaction cross section at 30 - 100\,MeV   & \cite{PhysRevC.104.024606}        & 0.278$\pm$ 0.035 & \\            % Starudubsky 1994
\hline
$^{208}$Pb+$^{208}$Pb at LHC  & \cite{PhysRevLett.131.202302}        & 0.217$\pm$ 0.058 & \\            % Giuliano Giacalone 2023
\hline
antiprotonic atoms                 & \cite{PhysRevC.76.014311}      & 0.16$\pm$(0.02)$_{stat}\pm$(0.04)$_{syst}$ &\\ % Klos 2007
                                   & \cite{PhysRevC.76.034316}      & 0.21$\pm$0.03 & \\                             % Wycech 2007
                                   & \cite{PhysRevC.76.034305}      & 0.20$\pm$(0.04)$_{exp}\pm$(0.04)$_{theo}$  & reanalysis of \cite{PhysRevC.76.014311} \\
\hline
pionic atoms                       & \cite{FRIEDMAN201246}      & 0.15$\pm$0.08 &\\ % Friedman 2012
                                   & \cite{FRIEDMAN201246}      & 0.14$\pm$0.10 & \\  % Friedmann
\hline

$\pi^+$ reaction cross section   & \cite{FRIEDMAN201246}      & 0.11$\pm$0.06 &  $^{nat}$Pb target\\  % Friedmann
\hline
strength of pigmy dipole resonance & \cite{PhysRevC.76.051603}      & 0.18$\pm$ 0.035& \\                            % Klimkiewicz 2007
                                   & \cite{PhysRevC.81.041301}      & 0.194 $\pm$ 0.024 &   \\                      % Carbone 2010
\hline
electric dipole polarizability     & \cite{PhysRevLett.107.062502}  & 0.156$^{+0.025}_{-0.021}$ &\\                 % Tamii 2011
by $\vec{p}$-scattering at 295\,MeV  & \cite{Tamii2014}               & \multicolumn{2}{l}{0.165$\pm$(0.009)$_{exp}\pm$(0.013)$_{theo}\pm$(0.021)$_{est}$ } \\ % Tammi 2014
                                   & \cite{PhysRevC.85.041302,PhysRevLett.107.062502}& 0.168$\pm$ 0.022 & reanalsis of \cite{PhysRevLett.107.062502}  \\  % Piekarewicz 2012 reanalyzed
\hline
giant dipole resonance; 120\,MeV $\alpha$-scattering& \cite{PhysRevLett.66.1287} & 0.19$\pm$ 0.09 & see \cite{KRASZNAHORKAY2004224} \\ % Krasnahorkay 1991
giant dipole resonance; 196\,MeV $\alpha$-scattering& \cite{CSATLOS2003C304,KRASZNAHORKAY2004224} & 0.12$\pm$ 0.07 & \\
\hline
anti-analog giant dipole resonance & \cite{Krasznahorkay_2013}      & 0.161$\pm$ 0.042 & \\                                  %Krasznahorkay 2013
                                   & \cite{10.1093/ptep/ptt038,PhysRevC.85.064606}     & 0.216$\pm$(0.046)$_{exp}\pm$(0.015)$_{theo}$  & \\ % Yasua 2013
\hline
coherent $\pi^0$ production & \cite{PhysRevLett.112.242502} & 0.15$\pm$0.03 (stat.) $^{+0.01}_{-0.03}$(sys.) & \\
\hline
parity violating e$^-$ scattering  & \cite{PhysRevLett.108.112502}  & 0.33$^{+0.16}_{-0.18}$ &\\                     % Abrahamyan 2012

parity violating e$^-$ scattering PREX 1+2  & \cite{PhysRevLett.126.172502}  & 0.283$\pm$0.071 &\\                     % Abrahamyan Adhikari 2021
\hline
tidal deformability from neutron star merger & \cite{PhysRevLett.120.172702} & $\leq$ 0.25 & analysis of \cite{PhysRevLett.119.161101short} \\
NICER & \cite{Riley_2019,Miller_2019} & $\leq$ 0.31 & analysis of \cite{PhysRevLett.126.172503}\\

\end{tabular}
\caption{The neutron skin thickness ${\Delta }$R$_{np}$ = R$_{rms}$(n)-R$_{rms}$(p) of $^{208}$Pb deduced by different experiments and analyses as shown in \autoref{fig:CaPb}.
}
\label{tab:nskinPb}
\end{ruledtabular}
\end{table*}
% --------------------- TABLE LL -----------------------------

%\include{PRC_Schupp_nskin_bib}


\end{document}